\documentclass{article}

\usepackage{arxiv}

\usepackage[utf8]{inputenc} 
\usepackage[T1]{fontenc}    
\usepackage{hyperref}       
\usepackage{url}            
\usepackage{booktabs}       
\usepackage{amsfonts}       
\usepackage{nicefrac}       
\usepackage{microtype}      
\usepackage{lipsum}		
\usepackage{graphicx}
\usepackage{natbib}
\usepackage{doi}
\usepackage{mathtools}
\usepackage{amsmath}
\usepackage{amssymb}
\usepackage{tikz}
\usetikzlibrary{positioning,shapes}
\usetikzlibrary{positioning, fit}
\usepackage{siunitx}
\usepackage{amsmath}
\usepackage{upgreek}

\title{Unveiling Causal Mediation Pathways in High-Dimensional Mixed Exposures: A Data-Adaptive Target Parameter Strategy}

\author{\href{https://orcid.org/0000-0002-5515-6307}{\includegraphics[scale=0.06]{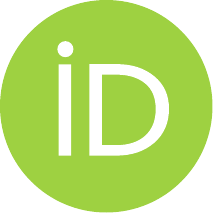}\hspace{1mm}David B. McCoy}\\
Department of Environmental Health Sciences\\
University of California, Berkeley\\
Berkeley, CA 94704 \\
\texttt{david\_mccoy@berkeley.edu} \\
\And
\href{https://orcid.org/0000-0002-3769-0127}{\includegraphics[scale=0.06]{orcid.pdf}\hspace{1mm}Alan E. Hubbard} \\
Department of Biostatistics\\
University of California, Berkeley\\
Berkeley, CA 94704 \\
\texttt{hubbard@berkeley.edu} \\
\And
\hspace{15mm}\href{https://orcid.org/0000-0003-1432-5511}{\includegraphics[scale=0.06]{orcid.pdf}\hspace{1mm}Mark J. van~der~Laan}\\
\hspace{15mm} Department of Biostatistics\\
\hspace{15mm} University of California, Berkeley\\
\hspace{15mm} Berkeley, CA 94704 \\
\hspace{15mm} \texttt{laan@berkeley.edu} \\
\And
\hspace{5mm}\href{https://orcid.org/0000-0003-4853-6130}{\includegraphics[scale=0.06]{orcid.pdf}\hspace{1mm}Alejandro Schuler} \\
\hspace{5mm} Department of Biostatistics\\
\hspace{5mm} University of California, Berkeley\\
\hspace{5mm} Berkeley, CA 94704 \\
\hspace{5mm} \texttt{alejandro.schuler@berkeley.edu} \\
}

\renewcommand{\shorttitle}{\textit{NOVAPathways}}

\begin{document}
\maketitle

\begin{abstract}
Mediation analysis in causal inference typically concentrates on one binary exposure, using deterministic interventions to split the average treatment effect into direct and indirect effects through a single mediator. Yet, real-world exposure scenarios often involve multiple continuous exposures impacting health outcomes through varied mediation pathways, which remain unknown \textit{a priori}. Addressing this complexity, we introduce NOVAPathways, a methodological framework that identifies exposure-mediation pathways and yields unbiased estimates of direct and indirect effects when intervening on these pathways. By pairing data-adaptive target parameters with stochastic interventions, we offer a semi-parametric approach for estimating causal effects in the context of high-dimensional, continuous, binary, and categorical exposures and mediators. In our proposed cross-validation procedure, we apply sequential semi-parametric regressions to a parameter-generating fold of the data, discovering exposure-mediation pathways. We then use stochastic interventions on these pathways in an estimation fold of the data to construct efficient estimators of natural direct and indirect effects using flexible machine learning techniques. Our estimator proves to be asymptotically linear under conditions necessitating $n^{-1/4}$‐consistency of nuisance function estimation. Simulation studies demonstrate the $\sqrt{n}$ consistency of our estimator when the exposure is quantized, whereas for truly continuous data, approximations in numerical integration prevent $\sqrt{n}$ consistency. Our NOVAPathways framework, part of the open-source SuperNOVA package in R, makes our proposed methodology for high-dimensional mediation analysis available to researchers, paving the way for the application of modified exposure policies which can delivery more informative statistical results for public policy. 
\end{abstract}



\section{Introduction}

Causal mediation analysis allows for the decomposition of an exposure's total effect on an outcome into direct and indirect pathways operating through an intermediate mediator or set of mediators. Identifying the pathways through which environmental mixtures impact health outcomes is crucial for corroborating causal inference of total effects and for developing effective public health policies. This information can help to strengthen causal inference by providing evidence for a plausible biological mechanism underlying the observed association between the exposure mixture and the outcome. Additionally, if several chemicals with similar structures are found to operate through the same mediating pathway, it may suggest that other chemicals with similar structures may have the same mediating effects. This type of inference is consistent with coherence used in the Bradford Hill criteria \cite{Fedak2015}. Such evidence can be used to strengthen regulations of unstudied chemicals which are structurally similar to chemicals which have been found to have both total effects and effects through certain biological pathways leading to disease. 

Mediation analysis can help in the development of targeted interventions in the context of environmental health by identifying specific pathways through which environmental exposures affect health outcomes. By determining the mediator variable(s) that link the exposure to the outcome, mediation analysis can help identify potential targets for intervention that may reduce the harmful effects of the exposure. For example, if mediation analysis identifies that inflammation is a key mediator between exposure to air pollution and cardiovascular disease, interventions that reduce inflammation, such as anti-inflammatory medications or dietary changes, may be targeted to reduce the harmful effects of air pollution on cardiovascular health in situations where air pollution cannot be immediately reduced. By identifying specific pathways, mediation analysis can help guide the development of targeted interventions that are more likely to be effective and efficient in reducing the harmful effects of environmental exposures.

Decomposing the total effects of a mixed exposures in environmental epidemiology presents unique challenges. Unlike single exposures, we do not know \textit{a priori} which specific exposures or sets of exposures act through which mediators to cause the outcome. There can be multiple such pathways, and using the same data to identify these pathways and estimate a target parameter given these pathways can lead to biased results due to overfitting to the sample data. That is, both the discovered pathway and effects for this pathway are overfit to the sample data and may not generalize to the population level. Additionally, it is possible that multiple exposures use the same pathways, and thus may interact through this pathway, such as multiple heavy metals interacting through epigenetic mediators, which can have more than additive effects through this pathway. This highlights the importance of developing methods that can identify and estimate the effects of mixed continuous-valued exposures on health outcomes through multiple mediators simultaneously while addressing issues of double-dipping and interactions between exposures. Currently, no such statistical methods exist to capture such complex exposure-outcome systems although, almost in all cases this is the system by which exposure leads to disease. 

Building upon the seminal work of Sewall Wright, who introduced path analysis in 1934 \cite{wright_1934}, researchers gained a foundation for exploring causal relationships among observed variables using path diagrams and standardized path coefficients. This approach enabled the decomposition of the total effect of one variable on another into direct and indirect effects via intermediary variables. In 1972, Arthur Goldberger \cite{goldberg_1972} further advanced the field by developing structural equation models (SEMs) for mediation analysis. By integrating path analysis with factor analysis, SEMs facilitated the modeling of intricate relationships between observed and unobserved (latent) variables. Goldberger's contribution linked path analysis to a more comprehensive statistical framework, providing enhanced precision in estimating causal effects while accounting for measurement error. Consequently, the scope and applicability of mediation analysis were significantly expanded. The initial development of path analysis and SEMs largely focused on parametric models, where assumptions about the distributional properties of the data and the functional form of relationships between variables were made. However, over time, researchers extended SEMs to include nonparametric and semiparametric approaches, allowing for more flexible modeling of relationships without strong distributional assumptions \cite{causal_inference_in_stats}.

In recent years, the field of causal inference has witnessed substantial advancements with the introduction of non-parametric structural equation models and directed acyclic graphs. These developments have facilitated the non-parametric estimation of causal effects and the evaluation of conditions that permit causal effect identification from data \cite{pearl_1995, robins1986, robins1992, rubin1974, robins2010}. While these novel approaches have addressed some limitations of traditional parametric structural equation models for mediation analysis, they also brought forth new challenges. Early non-parametric SEMs struggled with issues such as increased computational complexity; model identification, meaning that, in the absence of parametric assumptions, determining whether a non-parametric estimates are identifiable from the observed data is more challenging; sensitivity to choice of estimator; difficulty in assessing model fit and interpretability and limited available software. 

Non-parametric partitioning of the causal influence of a binary treatment into natural indirect and direct impacts began by employing the potential outcomes framework proposed by Robins and Greenland \cite{Robins1992IdentifiabilityAE}. The indirect impact measures the effect on the outcome variable via the mediator, while the direct impact measures the effect through all other pathways. Pearl \cite{pearl_2001} derived a similar effect partitioning utilizing non-parametric structural equation modeling. The identification of these natural (in)direct impacts depends on cross-world counterfactual independencies. Essentially, this means that we assume the outcomes of different imaginary scenarios, where intervention on the exposure and mediator, do not influence each other. The cross-world counterfactual independence assumption is not directly falsifiable from experimental data. This is because the assumption involves counterfactual variables that correspond to different hypothetical interventions, and we can only observe one intervention outcome in a single experiment. Therefore, the natural (in)direct impact is not identifiable in a randomized experiment, which means that even in in randomized experiments we cannot know if these estimated mediation effects actual exist at the population level for a deterministic intervention. 

These limitations arise because, in most causal inference research on mediation, deterministic interventions are studied, which assign fixed exposure values. Historically, binary exposures have been investigated for several reasons 1. interpretability: causal effects are easier to understand for binary exposures as they involve comparisons between two distinct groups or switching from one group to another; 2. estimation complexity: binary exposures often lead to simpler functional forms and estimation procedures, even in non-parametric settings; 3. identification: verifying assumptions for causal effects identification can be more straightforward for binary exposures; 4. potential outcomes framework: this framework is more intuitive for binary exposures, as there are only two potential outcomes for each individual. 

To avoid limitations of binary exposures while retaining interpretability and relaxed identification assumptions, stochastic interventions can be implemented. Stochastic interventions allow exposures to be a random variable after conditioning on baseline covariates. For example, in the context of air pollution exposure and cardiovascular outcomes, we can consider a stochastic shift intervention where air pollution exposure is reduced by an amount $\delta$ for each individual in the population. Therefore, this post-intervention distribution still depends on the originally observed air pollution levels. We then would estimate the impact under this post-intervention distribution and compare the average to the observed outcomes under observed air pollution exposures. Stochastic interventions offer analytical benefits over deterministic approaches by enabling the straightforward definition of causal effects for continuous exposures, providing an interpretation that is easily understood by those familiar with linear regression adjustment. Estimation of total effects for stochastic interventions has been explored in various studies, including methods for modified treatment policies and propensity score interventions for binary exposure distributions \cite{kennedy2018, diaz2012, stock1989, robins2004}. Nevertheless, these studies do not focus on decomposing the effects of stochastic interventions into direct and indirect effects, which was first investigated in \cite{diaz_2020}. 

In \cite{diaz_2020}, the authors introduce a decomposition of a stochastic intervention's effect into direct and indirect components. This approach identifies (in)direct effects without necessitating cross-world counterfactual independencies, producing experimentally testable scientific hypotheses that can be empirically tested by intervening on the mediator and exposure. The authors develop a one-step non-parametric estimator based on the efficient influence function, incorporating machine learning regression techniques, and provide $\sqrt{n}$-rate convergence and asymptotic linearity results. Importantly, the proposed method provides definition and estimation of non-parametric mediated effects for continuous exposures. However, in the software implementation of the proposed method, the authors employ a reparameterization of specific integrals as regressions and the authors treat the exposure as binary to reduce computation complexity by avoiding direct estimation of the probability density function (PDF) and estimating the probability mass function (PMF) instead. Likewise, restricting the software to a binary exposure also avoids numeric integration necessary for the estimator. While this approach enables the inclusion of multiple mediators, it necessitates a binary exposure to function effectively. This limitation motivates the work presented here. 

In many environmental epidemiology cases, it is crucial to understand the specific mediators through which particular exposures impact an outcome. Instead of reparameterizing an estimand to avoid high-dimensional density estimation or integrals, identifying individual mediators and estimating stochastic effects solely through these mediating pathways leads to deeper interpretation when dealing with multiple mediators. The random variables driving the outcome can be treated as  parameters, where these mediators are identified using one portion of the data, and direct/indirect effects are estimated for this mediator using another part of the data. Estimation becomes considerably more complex with multiple exposures, as the connections between exposures and mediators remain unknown. Thus, these paths must be discovered in the data, and mediation analysis employing stochastic interventions can then be estimated for these paths.

This study presents a methodological approach for estimating mediation effects in the presence of high-dimensional exposures and mediators. We employ a cross-validated framework, where in path-finding folds, a cross-validation process is used to identify the mediating paths through a series of semi-parametric regressions. With these paths established, we estimate the direct and indirect effects of a stochastic intervention on the exposure through the mediator, both identified in the path, in an estimation fold. Drawing on the efficient influence function from Diaz et al. \cite{diaz_2020}, we directly compute the integrals required for each component of the efficient influence function, rather than reparameterizing the estimates when the exposure is continuous. We also build in estimation for the case where the exposure is quantized, for example, into bins which represent quartiles. This approach enables the mediation of continuous/discrete exposures which are unknown \textit{a priori} through mediators which are also unknown \textit{a priori} and can take on multiple variable types.

The use of stochastic interventions in a semi-parametric framework provides a promising approach for estimating direct and indirect effects of exposure mixtures on health outcomes through mediation pathways. To our knowledge, no such methods exist in the causal inference literature which both makes available mediation for a continuous/discrete exposure and data-adaptive discovery of mediating paths. Our method proposed here is available for use in the SuperNOVA package in R which also estimates interaction and effect modification of a mixed exposure using stochastic interventions and data-adaptive target parameters. 

\section{The Estimation Problem}

Our mediation parameter of interest for a continuous exposure was first described in \cite{diaz_2020} and therefore, what follows in the our mediation framework for data-adaptively discovered mediation pathways is based on this previous work. That is, the notation, target parameter, identification and efficient influence function are all the same as in \cite{diaz_2020}; however we extend this method to work in the continuous case and data-adaptively identify exposure-mediator pathways. To make this current work more self-contained we review and explain with brevity parts of their estimator in order to describe our approach making estimates work in the fully continuous case of data-adaptively identified pathways.

We consider the causal inference problem involving a multivariate continuous, categorical, or binary exposure ($A$), a continuous, categorical, or binary outcome ($Y$), a multivariate continuous, categorical, or binary mediator ($Z$), and a vector of observed covariates ($W$) which are also a variety of data types. Let $O = (W, A, Z, Y)$ be a random variable with distribution $\mathbb{P}$. We denote the empirical distribution of a sample of $n$ independent and identically distributed observations $O_1, ..., O_n$ as $\mathbb{P}_n$. For any given function $f(o)$, we denote $\mathbb{P}f = \int f(o) d \mathbb{P}(o)$ and use $\mathbb{E}$ to represent expectations with respect to $\mathbb{P}$ averaging over all randomness. We assume $\mathbb{P}$ belongs to $\mathcal{M}$, the nonparametric statistical model comprising all continuous densities on $O$ with respect to a dominating measure $v$, with $p$ denoting the corresponding probability density function. We go through the framework first ignoring the data-adaptive selection of subsets of the $\{A,Z\}$. We then introduce the data-adaptive component which follows naturally. 

Our approach diverges from previous methods, focusing on data-adaptively identifying which sets of exposures ($\hat{A}$) impact which sets of mediators ($\hat{Z}$). This approach bypasses the need for estimating the high-dimensional joint impact of exposures through the mediators, an effort that often encounters the 'curse of dimensionality,' a phenomenon that complicates accurate modeling and prediction due to exponential increase in volume associated with adding extra dimensions in the exposure space. The discovered $\hat{A}$ and $\hat{Z}$ represent the "estimated" or selected subsets from the full set of $A$ and $Z$ variables.

Probability density functions and regression functions are represented as follows:
\begin{itemize}
\item $g(a | w)$: Represents the conditional probability density or mass function of $A$ given $W = w$.
\item $Q(a, z, w)$: Represents the expected outcome given the variables $A$, $Z$, and $W$.
\item $e(a | z, w)$: Represents the conditional density or probability mass function of $A$ given $(Z, W)$.
\item $q(z | a, w)$ and $r(z | w)$: Denote the conditional densities of $Z$.
\end{itemize}

To define our counterfactual variables, we use the following nonparametric structural equation model (NPSEM):

$$
W = f_W(U_W);
A = f_A(W, U_A);
Z = f_Z(W, A, U_Z);
Y = f_Y(W, A, Z, U_Y).
$$

This set of equations signifies a mechanistic model, grounded in nonparametric statistical methods, that is assumed to generate the observed data $O$. It incorporates several fundamental assumptions. First, an implicit temporal ordering is assumed, with Y occurring after $Z$, $A$, and $W$; $Z$ taking place after $A$ and $W$; and $A$ happening after $W$. Second, each variable (i.e., {$W$, $A$, $Z$, $Y$}) is assumed to be generated from the corresponding deterministic, yet unknown, function (i.e., {$f_W$, $f_A$, $f_Z$, $f_Y$}) of the observed variables that precede it temporally, as well as an exogenous variable, denoted by $U$. Each exogenous variable is assumed to encompass all unobserved causes of the corresponding observed variable.

In the context of nonparametric statistics, the independence assumptions on the exogenous variables $U = (U_W, U_A, U_Z, U_Y)$ necessary for identification will be addressed in the assumptions section. This approach allows the model to accommodate the complexities and nuances of the relationships between the variables without relying on specific functional forms or parametric assumptions.

Our causal effects of interest are characterized by hypothetical interventions on the NPSEM. In our situation, we focus on an intervention where the equation associated with $A$ is changed, and the exposure is drawn from a user-defined distribution $g_\delta(a | w)$. This distribution relies on $g$ (the conditional density under observed exposures) and is indexed by a user-specified parameter $\delta$. We assume that when $\delta = 0$, $g_\delta = g$. Let $A_\delta$ represent a draw from $g_\delta(a | w)$.

In our scenario, the distribution $g_\delta$ is given by $g(a - \delta | W)$, which indicates a shift of $\delta$ in the conditional density of $A$. This shift corresponds to a modified treatment policy aimed at reducing exposure by $\delta$. Essentially, the intervention involves removing the equation associated with $A$ and establishing the exposure as a hypothetical regime, $d(A, W)$. The regime $d$ depends on the natural exposure level $A$ (i.e., without any intervention) and covariates $W$. For instance, if $A$ denotes continuous exposures such as various air pollution factors (Carbon Monoxide, Lead, Nitrogen Oxides, Ozone, Particulate Matter, etc.) related to asthma incidence $Y$, we may be interested in investigating the expected asthma incidence if all individuals experienced a $\delta$-unit reduction in Lead exposure, while keeping other exposures and covariates unchanged. 

Assume that the distribution of $A$ given $W = w$ is supported within the interval $(l(w), u(w))$. In other words, the minimum pollution level for an individual with covariates $W = w$ is $l(w)$. We can then define a hypothetical post-intervention exposure, $A_\delta = d(A, W)$, as follows:

\begin{equation}
d(a, w) =
\begin{cases}
a - \delta & \text{if}\ a > l(w) + \delta \\
a & \text{if}\ a \leq l(w) + \delta
\end{cases}
\end{equation}

Here, $0 < \delta < u(w)$ is an arbitrary value provided by the user. This regime can be further refined by allowing $\delta$ to be a function of $w$, thereby enabling the researcher to specify a different change in pollution levels as a function of factors such as demographic characteristics or geographical location. This intervention was initially proposed by \cite{diaz_2012} and \cite{diaz2018} and \cite{haneuse2013}. 

We are interested in the population intervention effect (PIE) of $A$ on $Y$ using stochastic interventions. That is, given values for an exposure and mediator $(a, z)$, we examine the counterfactual outcome $Y(a, z) = f_Y(W, a, z, U_Y)$, the expected outcome if all individuals were exposed to these values for the exposure and mediator. We also examine the counterfactual mediator $Z(a) = f_Z(W, a, U_Z)$ or the expected value the mediator takes on given exposure $A = a$. The counterfactual $Y(a, z)$ represents the outcome in a hypothetical scenario where $(A, Z) = (a, z)$ is fixed for all individuals. We are interested in the contrast between the expected outcome given an intervention  $A_\delta$ which say, reduces exposure to pollution and the expected outcome under no intervention, the observed outcome under observed exposures. This looks like: 
$$
\psi(\delta) = \mathbb{E}\{Y(A_\delta) - Y\}.
$$

Drawing from causal inference literature on mediation, we know that since $A$ is a cause of $Z$, any intervention altering exposure to $A_\delta$ also affects the counterfactual mediator $Z(A_\delta)$. Owing to the consistency ensured by the NPSEM, we obtain $Y(A, Z) = Y$ and $Z(A)=Z$. In addition, from Pearl's \cite{pearl_2001} law of composition we can express $Y(A_\delta, Z(A_\delta)) = Y(A_\delta)$. In words, this means that the expectation of $Y$ under dual shift is implied by a shift in $A$ ignoring $Z$. Consequently, the PIE can be decomposed into a population intervention direct effect (PIDE) and a population intervention indirect effect (PIIE). The interepretation of these effects are the same as natural direct and indirect effects but are for a stochastic intervention rather than a deterministic intervention on $A$.

\begin{equation}
\psi(\delta) = \underbrace{\mathbb{E}\{Y(A_\delta, Z(A_\delta)) - Y(A_\delta, Z)\}}_{\text{PIIE}} + \underbrace{\mathbb{E}\{Y(A_\delta, Z) - Y(A, Z)\}}_{\text{PIDE}}.
\end{equation}

Essentially, the direct effect demonstrates the impact of an intervention that modifies the exposure distribution while maintaining the mediator distribution at the level it would have been without any intervention. On the other hand, the indirect effect quantifies the influence of an indirect intervention on the mediators, initiated by changing the exposure, while keeping the exposure intervention constant. 

\tikzset{state/.style={draw, rounded rectangle, minimum width=2cm, minimum height=1cm, text centered, text width=2cm, node distance=3cm}}
\begin{center}
    
\begin{tikzpicture}[->,shorten >=1pt,auto,semithick]

  \node[state] (A) at (0,0) {$A$};
  \node[state] (Z) at (3,0) {$Z$};
  \node[state] (Y) at (6,0) {$Y$};
  \node[state] (W) at (3,2) {$W$};
  
  \path (W) edge[->] node {} (A);
  \path (W) edge[->] node {} (Z);
  \path (W) edge[->] node {} (Y);
  \path (A) edge[->] node[above] {IE} (Z);
  \path (Z) edge[->] node[above] {IE} (Y);
  \path (A) edge[bend right=45,->] node[above] {DE} (Y);
  
  \node[align=center, below=of A] {Exposure \\ $a$};
  \node[align=center, below=of Z] {Mediator \\ $z$};
  \node[align=center, below=of Y] {Outcome \\ $y$};
  \node[align=center, above=0.5cm of W] {Covariates \\ $w$};
    
\end{tikzpicture}
\end{center}

Above is a simple directed acyclic graph (DAG) which illustrates the IE through the mediator $Z$ and DE which is the causal effect not through $Z$. For example, in a study investigating the effects of environmental exposure, such as air pollution, on respiratory health, the direct effect measures how changing pollution levels impact health outcomes, assuming the mediators (e.g., time spent outdoors) remain unchanged. The indirect effect, conversely, evaluates how health outcomes are influenced by changes in the mediators (e.g., reduced time spent outdoors) that result from modifying pollution levels, while the pollution intervention remains constant.

Above, $Y(A, Z)$ = $\mathbb{E}(Y)$, is simply estimated by the empirical mean in the sample. Moving forward, the optimality theory described in \cite{diaz_2020} which we review and estimators we present for the truly continuous exposure case focus on $\theta(\delta) = \mathbb{E}\{Y (A_\delta, Z)\}$. These two terms are then used in calculation of the direct effect. Because $\mathbb{E}\{Y (A_\delta, Z(A_\delta))\} = \mathbb{E}\{Y (A_\delta)$\}, which in words is simply the total effect in $Y$ after shifting $A$ ignoring $Z$. That is, if we were to construct an efficient estimator for a shift in $A$ ignoring $Z$ these estimates encapsulate the indirect effect $A$ has through $Z$ as in the total effect. Ivan Diaz and Mark van der Laan \cite{diaz_2012} first proposed estimators of the total effect of a stochastic shift intervention including inverse probability weighted, outcome regression, and doubly robust estimators based on the framework of targeted minimum loss-based estimation (TMLE) where in each case data adaptive machine learning can be used to estimate the relevant nuisance parameters. Call the total effect $\theta(\delta)_t$, which is the expected $Y$ given a shift in $A$ and includes the implied shift in $Z(A_\delta)$. Call $\mathbb{E}\{Y(A_\delta, Z) - Y(A, Z)\}$, $\theta(\delta)_d$, the direct effect or the effects of shift in $A$ keeping $Z$ fixed. Lastly,  $\mathbb{E}\{Y(A_\delta, Z(A_\delta)) - Y(A_\delta, Z)\}$, the effects of a shift in $Z$ due to a shift in $A$ keeping $A$ fixed we call $\theta(\delta)_i$. Then: 

$$
 \theta(\delta)_t = \theta(\delta)_d + \theta(\delta)_i
$$

Which means we can then estimate the indirect effect as: 

$$
 \theta(\delta)_i = \theta(\delta)_t - \theta(\delta)_d
$$

Which is simply estimating the indirect effect by subtracting the total effect from the direct effect, this provides us with the point estimate. We can do inference on this difference by utilizing work from \cite{diaz_2020} which provides an efficient estimator for $\theta(\delta)$ for the construction of the direct effect $\theta(\delta)_d$. We use TMLE or one-step estimators proposed from \cite{diaz_2012} to estimate $\theta(\delta)_t$ and we use the scalar delta method to estimate $\theta(\delta)_i$. Moving forward we describe $\theta(\delta)$, or $\mathbb{E}\{Y(A_\delta, Z)\}$, the average outcome under a shift in $A$ keeping $Z$ at natural values.

\subsection{Identification of the Causal Parameter}

We can evaluate the causal effect of our intervention by considering the counterfactual mean of the outcome under our stochastically modified intervention distribution. This target causal estimand is $Y(a,z)$, which is the counterfactual outcome we would observe when $\mathbb{P}{((A, Z) = (a, z))} = 1$. 

Our causal quantitiy is: 

$$ 
\theta(\delta) = \int y p_{Y(A_\delta, Z)}(y) \ dy
$$

\cite{diaz_2020} describe identification for this parameter and we briefly review here. We must assume that the data is generated by independent and identically distributed units, and that there is no unmeasured confounding, consistency, or interference (discussed in more detail in subsequent sections). Under these assumptions, $\theta({\delta})$ can be identified by a functional of the distribution of $O$: 



$$\theta({\delta}) = \int_{\mathcal{W}} \int_{\mathcal{A}} \int_{\mathcal{Z}} {Q}(a,z,w) g_{\delta}(a \mid w) r(z |  w)  q(w) \  dw\ da\ dz $$

Mechanically this is the outcome predictions from our $Q$ model integrated over density predictions from our $g$ model under $\delta$ shift integrated over our the conditional mediator and covariate density. 

Interpreting the statistical effects in our analysis as causal rests upon two assumptions: common support and conditional exchangeability (or ignorability). These are standard assumptions in causal inference that require consideration in mediation.

Common support, also known as positivity or overlap, is a fundamental assumption in causal inference that ensures that the distribution of the exposure of interest is well defined and supported by the data. For each individual in the population, there should be a non-zero probability of observing the shifted exposure value given their observed covariates. This assumption ensures that the exposure effect is identifiable and that causal inference can be validly conducted. In our case, positivity refers to the probability density of exposure being bounded away from zero or one after an exposure shift. We propose a method that data-adaptively finds a shift which does not lead to positivity violations (described later).

Conditional exchangeability, or ignorability, is related to the assumption made in \cite{vansteelandt2012}. In our context, it means that given the observed covariates, the distribution of the potential outcomes, $Y(a, z)$, is independent of the actual exposure, $A$, and mediator, $Z$, assignments. This assumption is akin to stating that we have adequately controlled for confounding.

Here, it's essential to note that we need conditional exchangeability both for the exposure-outcome and mediator-outcome relations. This implies that all confounders between the exposure $A$ and outcome $Y$, and between the mediator $Z$ and outcome $Y$, should be measured and properly adjusted for. If this assumption is violated—if there are unmeasured confounders—it can lead to biased effect estimates.

Consider the directed acyclic graph (DAG) below:

\begin{center}
\begin{tikzpicture}[
node distance=1.5cm,
every node/.style={draw, circle},
every edge/.style={draw, -latex}
]

\node (A) {A};
\node[right=of A] (Z) {Z};
\node[below=of Z] (V) {V};
\node[right=of Z] (Y) {Y};

\path (A) edge (Z);
\path (Z) edge (Y);
\path (V) edge (Z);
\path (A) edge[bend right] (Y);
\path (V) edge[bend left] (Y);

\end{tikzpicture}
\end{center}

This DAG illustrates the relations between the exposure $A$, mediator $Z$, confounder $V$, and outcome $Y$. Here, $V$ can be seen as a confounder that affects both $Z$ (the mediator) and $Y$ (the outcome). Conditioning on a collider $Z$ when there are unmeasured confounders ($V$), would open a pathway from $A$ to $Y(a, z)$, introducing bias into our estimates.

Additionally, the methods presented here cannot account for situations where the mediator-outcome confounder $V$ is affected by the exposure $A$. As this too opens up a backdoor path that would lead to bias \cite{diaz_2020}.

\subsection{Efficient Estimation of the Direct Effect}

In this section we focus on the efficiency theory for estimating $\theta(\delta)$ within the nonparametric model $\mathcal{M}$, with a focus on the efficient influence function (EIF) which was originally derived in \cite{diaz_2020}. Diaz and Hejazi offer a rigorous breakdown of the EIF for this part of the direct effect and we give a brief overview here to explain our approach for estimating each part of the EIF. The EIF is a fundamental concept in semi-parametric estimation theory. It plays a vital role in determining the asymptotic behavior of all regular and efficient estimators. In simpler terms, the EIF contains the information to predict how these estimators perform when the sample size approaches infinity. Calculating the EIF is crucial for constructing locally efficient estimators for $\theta(\delta)$. Locally efficient estimators are estimators that achieve the best possible asymptotic variance within a specified class of estimators under certain regularity conditions within a stated statisticla model. They are optimal in the sense that, asymptotically, they have the lowest variance among all unbiased estimators in their class. \cite{diaz_2020} derived the EIF for this problem: we briefly describe each part of the EIF here and describe how we estimate its components for the case where $A$ is a continuous/discrete exposure. The efficient influence function for $\theta(\delta)$ in the nonparametric model $\mathcal{M}$ for a modified treatment policy is $D^Y(o) + D^A(o) + D^{Z,W}(o) - \theta(\delta)$, where:

\begin{align*}
D^Y(o) &= \frac{g_\delta(a | w)}{e(a | z, w)}\{y - Q(a, z, w)\}, \\
D^{Z,W}(o) &= \int Q(a, z, w)g_\delta(a | w)\ da \\ 
D^{A}(o) &= \phi(a, w) - \int \phi(a, w)g(a | w) \ da
\end{align*}

Where: 

\begin{align*}
\phi(a, w) &= \int Q(d(a, w), z, w)r(z | w)\ dz \\
&= \mathbb{E}\left[\frac{g(A | W)}{e(A | Z, W)} Q(d(A, W), Z, W) \Big| A = a, W = w\right],
\end{align*}

Constructing an efficient estimator always involves estimating the EIF and so here we describe at a high level how we estimate each component in the rest of the article.

$D^Y$ describes the "weighting factor" in the EIF which adjusts the residuals of the outcome model ($Q$) based on the differences in exposure distributions between the intervention $g_{\delta}$ and the natural course of exposure $e$. We calculate this by directly constructing estimators for the conditional densities of $g$ and $e$. Likewise, $Q$ is simply an outcome regression model which is estimated using flexible machine learning. Therefore, in the case where $A$ is continuous, we use conditional density estimators to estimate the conditional density functions used in this nuisance parameter. When $A$ is discrete, we can also use an ensemble of multinomial regression estimators which provide the probability of exposure falling in each "bin". This probability mass function then replaces the probability density function used when $A$ is continuous.

$D^{Z,W}$ is expected outcome ($Q$) multiplied by the estimated probability density of the exposure under a shift by $\delta$, and integrating over all possible values of the exposure $a$. This takes into account the potential shift in the distribution of $w$ (which affects the exposure), to provide a more accurate prediction of the outcome $Y$. For this estimation we directly integrate the two functions over the exposure using Monte Carlo integration of the exposure variable over the exposure range. That is, exposure values $a$ are shifted until they meet the upper or lower bound in which case they simply take on the min or max value depending on the direction of $\delta$. In the case where $A$ is discrete, $g_\delta$ is simply the probability for the bin that corresponds to $a \pm \delta$ depending on the direction. For example, if $A$ is discretized into quartiles and $\delta$ is 1, then if $a$ is quartile 1, $g_\delta$ is the probability of quartile 2. In this case the integral is simply a weighted sum: 

$$
D^{Z,W}(o) = \sum_{a_k \in A} Q(a_k, z, w)g_\delta(a_k | w)
$$

For $D^{A}$ the first expression $\phi(a, w)$ can be calculated using either integration or regression. The first line of the expression uses integration to calculate this expected outcome by averaging the outcome model $Q$ over all possible values of $z$, weighted by the conditional density of $z$ given $w$, denoted as $r(z | w)$. The second line of the expression uses an alternative formulation to calculate the same expected outcome. It uses the conditional expectation formula to take the conditional expected value of $Q$ given $A=a$ and $W=w$, where the expectation is taken with respect to the conditional density of $A$ given $W$, denoted as $g(A | W)$, divided by the inverse of the conditional density of $A$ given $Z$ and $W$, denoted as $e(A | Z, W)$, which effectively regresses out the effect of $Z$ from $A$. Therefore, it possible to estimate $\phi(a, w)$ by either integrating or using pseudo-regression. We take both approaches to compare finite sample performance in both estimation approaches. For the integration approach, we directly estimate the conditional density of the mediator given covariates and use this function in the integration with $Q$ over $z$ using a Monte Carlo approach. Again if $A$ is discrete this looks like: 
$$
D^A(o) = \phi(a, w) - \sum_{a_k \in A} \phi(a_k, w) g(a_k | w)
$$

Because $\phi(a, w)$ is the integration of $Q$ and $r$ over $z$ and does not include $A$ as an outcome, it is still necessary to estimate the conditional density of $Z$ given $W$ even when the exposure is discrete. In this discrete exposure case, we use a double integration approach and pseudo regression approach.

\subsubsection{Monte Carlo Integration}

Monte Carlo (MC) integration is a numerical integration technique that uses random sampling to approximate the integral of a function over a given domain. In our case the range of exposures and/or mediators are the domains to integrate over. MC integration works by  first generating random points within the domain. Then, the function values are computed at these points. The  average of these values is then multiply by the volume of the domain. As the number of samples increases, the approximation converges to the true integral value.

In our case we are integrating the product of two density/regression estimators, for example in the case of, $D^{Z,W}(o) = \int Q(a, z, w)g_\delta(a | w) \, da$, MC integration can be more advantageous than quadrature methods for several reasons:

\begin{enumerate}
    \item Handling high-dimensional and non-linear functions: The product of fits using, for example, two Super Learners for $Q$ and $g$, may result in complex, non-linear, and high-dimensional functions. MC integration is well-suited for handling such functions, as it does not rely on any specific parametric assumptions or require the function to be smooth or continuous.
    \item Adaptability to irregular functions: MC integration is adaptive to irregularities in the function being integrated, making it a reasonable method for integrating the product of two flexible Super Learners fits, which can have irregular shapes across covariates. Quadrature methods, on the other hand, often rely on the function being smooth or continuous and may struggle with irregular functions.
    \item Scalability: MC integration is easily scalable to high dimensions, making it suitable for problems with a large number of covariates. Quadrature methods, in contrast, can suffer from the curse of dimensionality, where the number of required evaluation points grows exponentially with the dimensionality, leading to an intractable computational burden.
    \item Convergence properties: MC integration has desirable convergence properties, meaning that as the number of random samples increases, the accuracy of the approximation improves. This allows for obtaining more accurate estimates, even for complex and irregular functions.
    \item Ease of implementation: MC integration is relatively simple to implement and can be easily parallelized for efficient computation on modern hardware. Quadrature methods, on the other hand, can be more complex and challenging to implement, especially for high-dimensional and irregular functions.
\end{enumerate}

For these reasons, we use MC for estimating the necessary integrals of each nuisance function. MC integration is much faster than adaptive quadrature, especially in our case where we need to integrate these functions at every vector of covariates (for each observation). To ensure that the number of iterations is scaled by sample size the number of iterations used in the MC integration is set to four times sample size in this paper. 

\subsection{Estimation}

\subsubsection{Direct Effect}

\cite{diaz_2020} derive the efficient influence function $D_{\eta, \delta}$ to construct a robust and efficient estimator, which is defined as the solution to the estimating equation $P_n D_{\hat\eta, \delta}\hat{=} 0$ in $\theta$, given a preliminary estimator $\hat{\eta}$ of $\eta$. They advise utilizing cross-fitting in the estimation process to avoid entropy conditions of the initial estimators which we employ in our approach. To do this, the index set ${1, \dots, n}$ is randomly partitioned into $K$ equally sized estimation samples, $V_k$. For each $k$, the corresponding training sample $T_k$ is obtained by excluding $V_k$ from the index set. The estimator $\hat{\eta}_{T_k}$ is derived by training the prediction algorithm using only the data in $T_k$. The index of the validation set containing observation $i$ is denoted by $V_k(i)$.
The estimator is thus defined as:

\begin{equation}
    \hat{\theta}(\delta) = \frac{1}{n} \sum_{i=1}^{n} D_{\hat{\eta}_{k(i)}, \delta}(O_i) = \frac{1}{n} \sum_{i=1}^{n} \left[ D^Y_{\hat{\eta}_{k(i)}, \delta}(O_i) + D^A_{\hat{\eta}_{k(i)}, \delta}(O_i) + D^{Z, W}_{\hat{\eta}_{k(i)}, \delta}(O_i) \right]
\end{equation}

Effectively, the efficient estimator is the average of the cross-estimated sum of each nuisance parameter. Subtracting the mean from this sum of nuisance parameters then gives us the EIF for this shift parameter since the EIF is defined as $D^Y_{\eta,\delta}(o) + D^A_{\eta,\delta}(o) + D^Z_{\eta,\delta}(o) - \theta(\delta)$. 

When estimating $\theta(\delta)$ compared to the observed outcome, we employ the scalar delta method by subtracting the two efficient influence functions, resulting in an EIF for $\theta(\delta)_d$ that can be used for constructing confidence intervals and performing hypothesis testing. By subtracting the two EIFs and calculating the variance of the resulting EIF scaled by $n$ observations, we obtain the variance of $\theta(\delta)_d$, which is asymptotically Gaussian and centered around the true difference. Finally, we construct confidence intervals and conduct hypothesis testing using the standard error. This gives us our final point and variance estimates for the $\theta(\delta)_d$. 

\subsubsection{Indirect Effect}

\paragraph{One Mediator}

We employ one-step estimation or targeted maximum likelihood estimation (TMLE) to estimate the expected outcome of a shift in exposure $A$ without considering the mediator $Z$. TMLE solves the efficient influence function (EIF) and the delta method is used to estimate the total effect \cite{diaz_2012} by subtracting this EIF from the observed $Y$ EIF ($Y$ - $Q(a,w)$). By solving the EIF for the total effect parameter $\theta(\delta)_t$ using TMLE/one-step estimation and applying the delta method, we obtain the EIF for the indirect effect parameter $\theta(\delta)_i$ by subtracting $\theta(\delta)_d$ from $\theta(\delta)_t$, the same is done for the point estimates. Although we use different approaches for estimating $\theta(\delta)_t$ (TMLE) and $\theta(\delta)$ (estimating equations), both result in efficient estimators. According to the central limit theorem, the distribution of each estimator is Gaussian and centered at the true value. We can compute the estimate of the variance $\sigma_n^2$, allowing for Wald-style confidence intervals to be computed at a coverage level of $(1 - \alpha)$ as $\psi_n \pm z{(1 - \alpha/2)} \cdot \sigma_n / \sqrt{n}$.

\paragraph{Many Mediators}

In research situations where multiple mediators are measured, we need to adjust the above described methodology in order to isolate the indirect effect for a given pathway. A simple subtraction of the direct effect from the total effect to derive the indirect effect when multiple potential mediators are present would yield an oversimplified estimation. This approach would instead estimate the collective indirect effect through all potential mediators. This would not provide the specific indirect effect attributable to the pathways of interest. To delineate the specific indirect effect through the mediator of interest, we adopt a slightly different approach. When we estimate the total effect, we adjust for all the other mediators but not the mediator of interest in the model, symbolized as $\mathbb{E}[Y|A, Z_{\neg i}, W]$ where $Z_{\neg i}$ represents all mediators other than the mediator of interest. This enables us to isolate the total effect of $A$ on $Y$ with respect to the particular $A-Z$ pathway under investigation. The rest of the estimation procedure is the same where we subtract this total effect point estimate and EIF from the direct effect to get the indirect effect estimates.

\section{Finding Mediating Pathways}
\subsection{Mixed Exposures and Mediators}

Up to this point, we have focused on fixed exposure $A$ and mediator $Z$, showing the efficient influence functions (EIFs) necessary for estimating the natural (in)direct effects. However, in scenarios involving mixed exposure and multiple potential mediating paths, the most important exposure-mediator ($A-Z$) paths among a potentially high-dimensional set are typically unknown. 

Consider a hypothetical situation where five exposures represent different pesticides ($A_{1-5}$), and five measured variables potentially mediate the effects of these pesticides ($Z_{1-5}$), representing neurotoxicity, endocrine disruption, oxidative stress, immune system, and DNA damage. In this scenario, let us imagine that the effects of $A_1$ and $A_2$ are mediated through $Z_1$ and $Z_2$ respectively, while $A_3$ shows no measured indirect effects, and $A_4$ and $A_5$ have no impact on the outcome. Additionally, $Z_3-Z_5$ do not mediate any measured exposures. A directed acyclic graph (DAG) illustrating this situation is presented below.

\tikzset{state/.style={draw, rounded rectangle, minimum width=2cm, minimum height=1cm, text centered, text width=2cm, node distance=3cm}}
\begin{center}

\begin{tikzpicture}[->,shorten >=1pt,auto,semithick]

\node[state] (A1) at (0,0) {$A_1$};
\node[state] (Z1) at (3,0) {$Z_1$};
\node[state] (Y) at (6,1) {$Y$};
\node[state] (Z2) at (3,2) {$Z_2$};
\node[state] (A2) at (0,2) {$A_2$};
\node[state] (A3) at (0,-2) {$A_3$};
\node[state] (W) at (3,5) {$W$};

\path (W) edge[bend right=95,->] node {} (A1);
\path (W) edge[bend right=95,->] node {} (A3);
\path (W) edge[bend right=45,->] node {} (A2);
\path (W) edge[bend right=45,->] node {} (Z1);
\path (W) edge[bend left=45,->] node {} (Y);
\path (W) edge[->] node {} (Z2);
\path (A1) edge[->] node[above] {IE} (Z1);
\path (Z1) edge[->] node[below] {IE} (Y);
\path (A1) edge[bend right=45,->] node[below] {DE} (Y);
\path (A2) edge[->] node[above] {IE} (Z2);
\path (Z2) edge[->] node[above] {IE} (Y);
\path (A2) edge[bend left=45,->] node[above] {DE} (Y);
\path (A3) edge[bend right=45,->] node[below] {DE} (Y);

\end{tikzpicture}
\end{center}

Scenarios featuring mixed exposures and multiple mediating pathways are not uncommon in real-world contexts. For instance, agricultural workers may encounter multiple pesticides, chemicals, and environmental factors, each acting through different mediating pathways to exert health effects. Similarly, industrial employees can be exposed to various chemicals, urban residents to diverse air pollutants, and individuals practicing unhealthy lifestyle habits to the risk of chronic diseases. Even the spread of infectious diseases and climate change can involve a complex interplay of multiple exposures and mediating pathways.

In all these instances, understanding the complex interplay between various exposures and mediating pathways is crucial. However, since the $A-Z$ pathways are not known \textit{a priori}, and continuously testing different exposure-mediator pathways could lead to type 1 error, a data-driven approach is essential.

\subsection{Basis Function Estimators for Pathway Discovery}

Uncovering mediating pathways in our data requires a non-parametric method, as not only are pathways not known \textit{a priori} but also the functional forms underlying their relationships are not known as well. We leverage a series of discrete Super Learners—best fitting flexible estimators selected from a library of candidate estimators—for this task. These constituent learners used in the Super Learner construct non-linear models through linear combinations of basis spline terms and their tensor products, rendering them ideal for the task at hand. 

In the most flexible setting, we form indicator variables for each predictor. These variables denote if a predictor $X$ is less than or equal to a specific value $x_s$, this approach can be extended for combinations of predictors, like $X_1, X_2$. Consequently, a function of our outcome distribution can be represented as:

$$\psi_{\beta} = \beta_0 + \sum_{s\subset \{1,2,...,p\}}\sum_{i=1}^{n} \beta_{s,i} \phi_{s,i},
    \text{ where } \phi_{s,i} = I({X}_{i,s} \leq x_s), A \in \mathbb{R}^p$$

Here, $s$ denotes indices of subsets of the $X$.

This estimator is known as the highly adaptive lasso (HAL) estimator \cite{hal_paper}. Its unique attribute is its theoretically proven $n^{-1/4}$ convergence, a necessary condition for the $\sqrt{n}$ rate conditions to hold for our estimator and for convergence to selection of basis functions of true pathways for the underlying yet unknown DGP. However, the HAL estimator is not scalable in high dimensions. Therefore, the estimators employed in NOVAPathways that return tensor products of arbitrary order and approximate this more exhaustive approach include the earth \cite{earth}, polySpline \cite{polspline}, and hal9001 (under restrinctions) \cite{hal9001} packages. Each method utilizes a linear combination of basis functions to estimate the conditional outcome, allowing us to extract variable sets used in these basis functions as our data-adaptively identified variable sets.

Our process to construct pathways includes:

\begin{enumerate}
    \item Fitting $\mathbb{E}(Z|A,W)$ as $\beta_0 + \sum[\beta_s \cdot h(A, W)_s]$. Here, $\beta_0$ is the intercept, $\beta_s$ are the coefficients, $h(A, W)_s$ are the basis functions involving $A$ and $W$, and the sum is over all basis functions in the model.
    \item Extracting basis functions for $A$ with non-zero coefficients.
    \item Fitting $\mathbb{E}(Y|A,Z,W) = \beta_0 + \sum[\beta_s \cdot g(A, Z, W)_s]$.
    \item Matching A to Z pathways: we align the basis functions involving $A$ from the first model ($\mathbb{E}(Z|A,W)$) with the basis functions involving $Z$ from the second model ($\mathbb{E}(Y|A,Z,W)$), if used, to identify the $A-Z$ pathways. Pathways are also $AZ$ basis functions used directly in the second model.
\end{enumerate}

This stepwise approach is necessary. In cases where the effects of $A$ go entirely through $Z$, or when effects that don't pass through $Z$ are negligible for the model fit, the second model will only contain basis functions for $Z$. As such, the first model is required to illuminate the underlying $A$ driver, thereby establishing the pathway connection. In summary, this approach non-parametrically identifies mediating pathways in a mixed exposure scenario.

\subsection{Non-Parametric Analysis of Variance for Identifying "Important" Pathways}

Upon identification of the optimal basis spline estimator for each sequential regression segment, we use an ANOVA-like decomposition of basis functions to rank the "important" variable sets employed by each algorithm; thereby filtering to the most important pathways. This selection process becomes critical in high-dimensional $A$ and $Z$ scenarios, where the possible paths multiply, and the goal is to discern the most influential pathways on $Y$ adaptively. We apply a variant of ANOVA, generalized for large-scale, non-parametric models.

In this context, we partition the response variable's variance based on the contributions from distinct basis factors. For multivariate adaptive regression models, the variance is decomposed into the individual contributions of linear basis functions. For highly adaptive lasso models, zero-order basis functions (exposure-covariate indicators) make these contributions. In both cases, the F-statistic is calculated using the traditional ANOVA formula, albeit with modifications to accommodate the non-linear model concerning the original covariates.

The response variable's variance is split into two: variance explained by the linear combination of basis functions and the residual variance left unexplained by the model. The F-statistic represents the ratio of explained to residual variance, adjusted for degrees of freedom. The F-statistic is computed for each basis using the standard formula, presuming a linear relationship between the response variable and basis functions, though the basis functions themselves need not be linear in the original covariates. 

Once we've computed F-statistics for each basis function, we have a measure of each basis function's contribution to the explained variance in the response variable. However, these basis functions represent transformations (which may or may not be linear) of the original variables. Hence, we're interested in getting a measure of the importance of each variable, not just the individual basis functions.

To aggregate these F-statistics to the variable level, we need to map each basis function back to the original variables it was derived from. We do this using the naming conventions of the basis functions, which contain the names of the variables they were derived from. This allows us to identify which F-statistics belong to which variables.

Once this mapping is complete, we have a collection of F-statistics for each variable, with each statistic representing a different basis function of that variable. To aggregate these statistics, we take their sum. The sum provides a measure of the total contribution of all basis functions of a variable to the explained variance in the response variable. In other words, it gives us a measure of the overall importance of that variable.

Finally, we rank the variables according to these sums of F-statistics, which we can then use to filter variables in subsequent analyses. It's important to note that this approach assumes that the F-statistics of basis functions of a variable can be meaningfully added together. This assumption holds true if the basis functions are orthogonal (i.e., uncorrelated), as is the case with splines. However, it may not hold if the basis functions are correlated, which might be the case with other types of basis functions.

We then rank variable sets based on the computed F-statistics, and subsets can be decided based on the F-statistic quantile to yield a concise variable list. The resultant list contains variable sets that meet the F-statistic threshold. This procedure is applied to both $\mathbb{E}(Z|A,W)$ and $\mathbb{E}(Y|A,Z,W)$ models, to filter $A$ based on the F-statistics driving each mediator, and to filter $Z$, $A$, and $A-Z$ basis functions, respectively. This variable set process is implemented within a V-fold cross-validation framework using data, which we discuss in the subsequent section.

It's worth noting that our proposed methodology operates on the principle of heuristics, aiming for an approach that's both computationally practical and effective. It's not designed to achieve theoretical optimality but rather to robustly identify potential pathways, that are part of the data-adaptive target parameter. Theoretical rigor is still maintained during the estimation step.  While other methods could be employed, our approach offers a blend of simplicity, speed, and suitability for the task at hand by using basis-function estimators in the two step process that are both flexible but also interpretable, which allows us to construct the pathways. For example, methods could be employed such as using exposure or exposure-mediator sets used in the branches of a best fitting decision tree \cite{McCoy2023} to identify potential pathways.

\section{Cross-Estimation}

Ensuring the estimators of our mediation target parameters meet the requisite complexity conditions, such as smoothness (differentiability) and entropy small enough to satisfy the Donsker conditions, can be challenging in high-dimensional settings $(p > n)$ that necessitate complex/adaptive ML methods. Although verifying entropy conditions is feasible for certain machine learning techniques like lasso, it becomes notably difficult with methods involving cross-validation or hybrid models, such as Super Learner.

To address this, we employ a strategy of sample splitting. This approach separates the data into two independent sets: one for estimating the nuisance functions and the other for constructing the mediation parameters. Originally proposed by Bickel and later refined by Schick, this strategy has been extended to k-fold cross-validation, allowing for the average mediation estimates from different data partitions to be employed.

Sample splitting allows us to handle the more complex task of identifying mediating pathways within high-dimensional data. Typically, we lack prior knowledge of these pathways amidst a diverse mixture of exposures and mediators, necessitating data-adaptive identification methods. The separate data partitions help ensure that pathway discovery and estimation of direct and indirect effects are not overfit to the sample data, thus avoiding the statistical pitfall of double-dipping, which is akin to multiple testing issues.

This process of identification has been termed "dredging with dignity" in the literature \cite{hubbard2016}, recognizing the necessity of exploring the vast array of potential paths in a principled manner. Just as the analyst might be tempted to cherry-pick interesting results from multiple testing, so too can the analyst fall into the trap of selecting intriguing pathways from the same dataset. This separate sample approach steers clear of that, offering a way to explore high-dimensional pathways responsibly.

\subsection{K-fold Cross-Validation}

K-fold cross-validation is a technique that divides our observations, indexed from 1 to $n$, into $K$ equally sized subgroups. For each $k$, an estimation sample $P_k$ is defined as the $k$-th subgroup of size $\frac{n}{K}$, while the complement of $P_k$, denoted as $P_{n_{-k}}$, serves as the parameter-generating sample. Using $P_{n_{-k}}$, we identify mediation pathways in the exposure-mediator space by employing basis functions from the best-fitting b-spline estimators. In each fold, we have nuisance estimators for every component of the EIF. With these  mediation pathways fixed we then train nuisance parameter estimators on the same $P_{n_{-k}}$ samples, which are essential for solving the EIF and providing asymptotically unbiased estimators.

The process is carried out in a round-robin manner. For $K=10$, we obtain 10 (possibly different) pathways, outcome estimators $Q_k$, and density estimators $g_k$, $e_k$, and $r_k$, which are used to construct nuisance parameters that comprise $D^Y(o)$, $D^A(o)$, and $D^{Z,W}(o)$. To estimate a pooled $\theta(\delta)$ using the full data, we stack the estimation-sample estimates for each nuisance parameter across the folds. We then calculate the sum and average across the folds to obtain our point estimate and subtract this average from the summed nuisance parameters to obtain the EIF for the full data, yielding our pooled $\theta(\delta)$ estimate. The variance is then calculated by pooling the pooled EIFs. The NDE parameter is obtained by subtracting the pooled $\theta(\delta)$ from the full data mean outcome, and the delta method is applied to the pooled EIFs to obtain the EIF for the pooled NDE $(\theta(\delta)_d)$, which is used to derive confidence intervals (CIs).

A similar procedure is employed for the total effect. For the total effect we are using TMLE or one-step estimation. For TMLE, we stack initial estimates and clever covariates across all folds and perform a fluctuation step across the full initial estimates and clever covariate estimates to obtain our estimate $\epsilon$. We then update the counterfactuals across all folds using the $\epsilon$ values. The updated conditional means, counterfactuals, and clever covariates are employed to solve the EIF across the entire sample for the shift in $A$, ignoring $Z$. The delta method is applied to subtract the EIF for a shift in $A$, ignoring $Z$, from the EIF of the observed $Y$ to obtain the EIF for the total effect $(\theta(\delta)_t)$, and the same process is used to derive the point estimate for the total effect. The delta method is again used to estimate the pooled NIE $(\theta(\delta)_i)$.

\begin{figure}[!h]
  \hspace*{-2cm}\includegraphics[scale = 0.6]{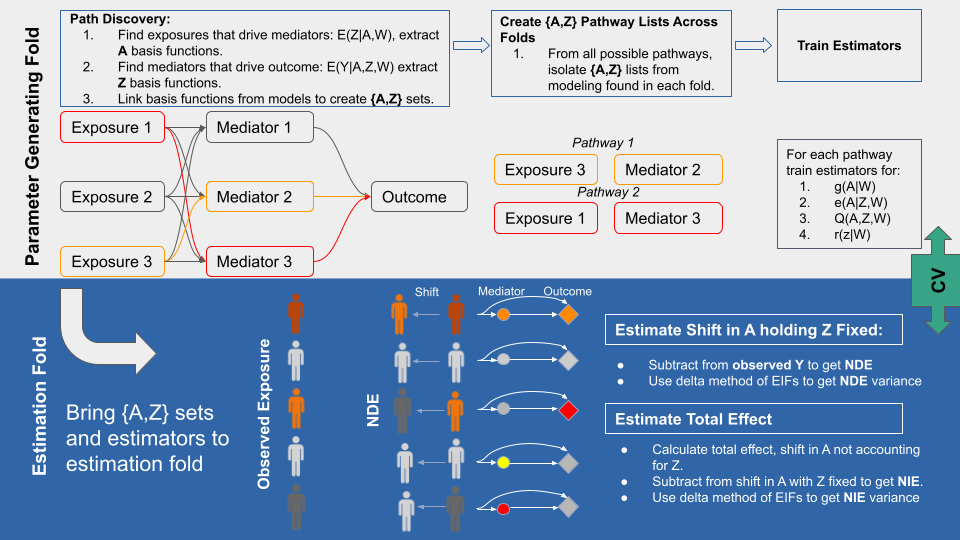}
  \caption{Schematic of Operations in the Parameter Generating and Estimation Folds in the NOVAPathways Procedure}
  \label{fig:NOVAPathways_schematic}
\end{figure}

In addition to the pooled estimates, we report k-fold specific estimates of the in(direct) effects and fold-specific variance estimates for these target parameters using the fold-specific IC. This is important because if the mediation pathway identified in each fold varies significantly, the pooled estimates can be challenging to interpret (if the same pathway is not found across all folds). By providing both k-fold specific and pooled results, users can assess the robustness of the pooled result across the folds. To visualize the algorithm and what is happening in the parameter generating and estimation folds, we provide a schematic in \textbf{Figure \ref{fig:NOVAPathways_schematic}}

\subsection{Pooled Estimates under Data-Adaptive Delta}

Stochastic interventions, particularly those involving significant shifts in exposure, can be susceptible to positivity violations, leading to biased and increased variance in exposure effect estimation. This happens if the exposure shift is so substantial that some subgroups have zero probability of receiving a specific exposure level. This challenge persists even when utilizing an efficient estimator like TMLE. 

To mitigate this, we can employ a data-adaptive approach to adjust the exposure shift magnitude, $\delta$. When the exposure is continuous, we modify $\delta$ within the parameter-generating sample to meet specific positivity criteria, which helps limit positivity violations. However, when exposure is quantized, a delta of one, signifying an increase in quantiles, is the minimum and most interpretable $\delta$.

Consider $H(a_\delta,w)_i$, the probability density ratio for observation $i$ upon an exposure shift of $\delta$. We aim to ensure that all observations have a ratio below a specific threshold $\lambda$. To do this, we iteratively decrease $\delta$ by a small amount, $\epsilon$, until $H(a_\delta,w)_i < \lambda$ for all observations $i$:

$$\forall i, \;  H(a_\delta,w) = \frac{g_{n_{-k}}(a_{n_{-k}} - \delta \mid w)}{g_{n_{-k}}(a_{n_{-k}} \mid w)} \le \lambda$$

Here, $\lambda$ is a preset threshold, and $\delta$ is reduced until all clever covariate density ratios fall below $\lambda$. By default, in our SuperNOVA package, $\lambda$ is set to 50, and $\epsilon$ to 10\% of $\delta$. This means that if any predicted conditional probabilities exceed the probability under observed exposure by a factor of 50, we reduce $\delta$.

Finally, we account for data-adaptive $\delta$ during the pooling process. If $\delta$ is constant, the pooling is simply an average of the estimates across folds. However, for a data-adaptive $\delta$, we average $\delta$ and pair it with the average estimates and the pooled variance calculations described previously.

\subsection{Interpreting Shifts when Exposure is Discretized}

In the case where the exposure variable is discretized into quantiles, we can still interpret the results in a continuous context. If we let $A_{\text{min}}$ and $A_{\text{max}}$ denote the minimum and maximum values of the continuous exposure, respectively, and $n_{\text{bins}}$ denote the number of quantiles, then each quantile represents an interval of size $\frac{A_{\text{max}} - A_{\text{min}}}{n_{\text{bins}}}$ on the continuous scale. 

\begin{equation}
A_{\text{quantile}} = A_{\text{min}} + \left(\frac{A_{\text{max}} - A_{\text{min}}}{n_{\text{bins}}}\right) \cdot (q - 1)
\end{equation}

Here, $q$ denotes the quantile rank, which ranges from 1 (for the smallest values of the exposure) to $n_{\text{bins}}$ (for the largest values of the exposure). Each value of $A_{\text{quantile}}$ represents the lower bound of the interval on the continuous scale that corresponds to that quantile.

For example, if we have an exposure variable ranging from 0 to 10, and we discretize it into 5 quantiles, then each quantile represents an interval of size $\frac{10 - 0}{5} = 2$ on the continuous scale. Thus, the first quantile represents the interval from 0 to 2, the second quantile represents the interval from 2 to 4, and so on. Despite using a discretized version of the exposure in the analysis, the interpretation can still be related back to the original continuous exposure scale. In this way, if the discretized approach is preferable, a pseudo-continuous interpretation is still possible.

\section{Simulations}
In this section, we demonstrate using simulations that our approach identifies the correct mediating pathways in a complex mixture of exposures and mediators and correctly estimates the natural direct, indirect and total effects for a given pathways using stochastic interventions.

\subsection{Data-Generating Processes}

We first construct a simple data-generating process (DGP) where $Y$ is generated from a linear combination of an exposure and mediator. In this DGP we measure the asymptotical behavior of the in(direct) effect estimators keeping the pathway fixed (not data-adaptively discovering the pathway). We do simulations for both continuous and discrete exposures to investigate the behavior of the estimator using numeric integration vs. simple weighted sums. In the second DGP, we generate multiple pathways to the outcome from multiple exposures and measure the estimators performance in data-adaptively identifying the correct pathways.

\subsubsection{Simple Mediation Simulation}

This DGP has the following characteristics, $O = W,A,Z,Y$. We call this the "DGP 1" moving forward which we use to investigate the asymptotic behavior of our estimates. The data-generating process involved the following steps:

\begin{enumerate}
    \item Baseline covariates:
    \begin{itemize}
        \item $W_1 \sim \mathcal{N}(20, 2^2)$: generated from a normal distribution with mean 20 and standard deviation 2.
        \item $W_2, w_3 \sim \text{Binomial}(1, 0.5)$: generated from binomial distributions with size 1 and probability 0.5.
        \item $W_4 \sim \mathcal{N}(30, 3^2)$: generated from a normal distribution with mean 30 and standard deviation 3.
        \item $W_5 \sim \text{Poisson}(1.2)$: generated from a Poisson distribution with rate 1.2.
    \end{itemize}

    \item The exposure $A$ was generated from a normal distribution, conditional on the covariate $W_1$:
    $$A \sim \mathcal{N}(1 + 0.5 W_1, 1^2)$$
    \item The exposure $A$ was shifted by an amount $\delta$ (in this example $\delta$ = 1), producing $A_{\delta}$:
    $$A_{\delta} = A + \delta$$
    \item The mediator $Z$ was generated from a normal distribution, conditional on the exposure $A$ and the covariate $W_1$:
    $$Z \sim \mathcal{N}(2 \cdot A + W_1, 1^2)$$
    \item The mediator $Z$ was also shifted given a shift in $A$, producing $Z_{A_\delta}$:
    $$Z_{A_\delta} \sim \mathcal{N}(2 \cdot A_\delta + W_1, 1^2)$$
    \item The outcome $Y$, $Y_{A_\delta}$ ($Y$ given a shift in only $A$), and $Y_{A_\delta, Z_{A_{\delta}}}$ ($Y$ given a shift in $A$ and $Z$)   was generated as a linear function of the exposure $A$ and the mediator $Z$:
    $$Y = 10 \cdot Z + 40 \cdot A + \epsilon$$
    $$Y_{A_\delta} = 10 \cdot Z + 40 \cdot A_\delta + \epsilon$$
    $$Y_{A_\delta, Z_{A_{\delta}}} = 10 \cdot Z_{A_\delta} + 40 \cdot A_\delta + \epsilon$$
    
\end{enumerate}

We use this simulation to test NOVAPathways estimation of the total, direct and  indirect effects. Our approach was to keep things relatively straightforward, keeping the DGP a linear process to test the asymptotic behavior of the estimator when the functional forms are correctly specified (GLMs are included in each Super Learner that model the true underlying function).

\subsubsection{Complicated Mediation Simulation}

We now want to create a more complicated scenario where there are many correlated exposures and some go through mediators to drive the outcome. In this simulation, we want to test NOVAPathways in discovering the correct paths. This data-generating process (DGP) has the following characteristics, $(O = (W,A,Z,Y))$, we call this moving forward "DGP 2". The exposures $A = (A_1, A_2, A_3, A_4, A_5)$ are generated and have potential indirect (through $Z = (Z_1, Z_2, Z_3, Z_4, Z_5)$) effects on $Y$. Even though there are a total of 25 possible mediating paths due to 5 exposures and 5 mediating variables, only the exposures $A_1, A_2$ have actual direct and indirect (through $Z_1, Z_2$ respectively) effects on $Y$. Our goal is to test the proportion of times across the simulation that the correct paths among the 25 potential ones are discovered. The data-generating process involved the following steps:

\begin{enumerate}
\item Baseline covariates:
\begin{itemize}
\item $W_1 \sim \mathcal{N}(20, 2^2)$: generated from a normal distribution with mean 20 and standard deviation 2.
\item $W_2, W_3 \sim \text{Binomial}(1, 0.5)$: generated from binomial distributions with size 1 and probability 0.5.
\item $W_4 \sim \mathcal{N}(30, 3^2)$: generated from a normal distribution with mean 30 and standard deviation 3.
\item $W_5 \sim \text{Poisson}(1.2)$: generated from a Poisson distribution with rate 1.2.
\end{itemize}
\item Five exposure variables $A = (A_1, A_2, A_3, A_4, A_5)$ are generated from a multivariate normal distribution, conditional on the covariates $W$:
\begin{itemize}
    \item $A_1 \sim \mathcal{N}(1 + 0.5 \cdot W_1, \Sigma)$
    \item $A_2 \sim \mathcal{N}(2 \cdot W_2 \cdot W_3, \Sigma)$
    \item $A_3 \sim \mathcal{N}(1.5 \cdot W_4 / 20 \cdot W_1 / 3, \Sigma)$
    \item $A_4 \sim \mathcal{N}(3 \cdot W_4 / 2 \cdot W_2 / 3, \Sigma)$
    \item $A_5 \sim \mathcal{N}(2 \cdot W_5, \Sigma)$
\end{itemize}
The exposures are correlated as per the following correlation matrix, $\Sigma$, which represents common scenarios in air pollution where particulate matter and gaseous pollutants show high intra-group correlation but lower inter-group correlation:

$$
\Sigma = \begin{bmatrix}
1 & 0.8 & 0.3 & 0.3 & 0.2 \\
0.8 & 1 & 0.3 & 0.3 & 0.2 \\
0.3 & 0.3 & 1 & 0.8 & 0.2 \\
0.3 & 0.3 & 0.8 & 1 & 0.2 \\
0.2 & 0.2 & 0.2 & 0.2 & 1 \\
\end{bmatrix}
$$

\item Five mediators $Z = (Z_1, Z_2, Z_3, Z_4, Z_5)$ are generated from normal distributions, conditional on the exposures $A$ and covariates $W$:
\begin{itemize}
    \item $Z_1 \sim \mathcal{N}(2 \cdot A_1 + W_1, 1^2)$
    \item $Z_2 \sim \mathcal{N}(2 \cdot A_2 + W_2, 1^2)$
    \item $Z_3 \sim \mathcal{N}(5 \cdot A_3 \cdot A_4 + W_3, 1^2)$
    \item $Z_4 \sim \mathcal{N}(3 \cdot A_4 \cdot W_4, 1^2)$
    \item $Z_5 \sim \mathcal{N}(4 \cdot A_5 \cdot W_5, 1^2)$
\end{itemize}

\item The outcome $Y$ is generated as a linear function of $Z_1$, $A_1$, $W_3$, $A_2$, and $Z_2$:
    $$Y = 10 \cdot Z_1 + 40 \cdot A_1 + 15 \cdot W_3 - 6 \cdot A_2 + 7 \cdot Z_2$$
\end{enumerate}

\subsubsection{Calculating Ground-Truth}

We numerically approximated the natural direct effect (NDE), natural indirect effect (NIE), and total effect (ATE) of the exposure $A$ on the outcome $Y$ to high precision using 100000 samples from our DGP. In our DGP which assesses estimation (simple DGP), $\delta$ is equal to 1.

\begin{enumerate}
    \item The NDE was calculated as the mean difference in $Y$ when shifting the exposure $A$ while keeping the mediator $Z$ constant:
    $$\text{NDE} = \mathbb{E}(Y_{A_\delta, Z} - Y)$$
    \item The NIE was calculated as the mean difference in $Y$ when shifting both the exposure $A$ and the mediator $Z$:
    $$\text{NIE} = \mathbb{E}(Y_{A_\delta, Z_{A_\delta}} - Y_{A_\delta})$$
    \item The ATE was calculated as the sum of NDE and NIE:
    $$\text{ATE} = \text{NDE} + \text{NIE}$$
\end{enumerate}

Additionally, we conducted the same analysis using a discrete exposure that has been split into quantiles (10) after step 2. and compute the quantile-based NDE, NIE, and total effect estimates. 

\subsection{Evaluating Performance}

We assessed the asymptotic convergence to the true exposure relationships used in the DGP, as well as the convergence to the true in(direct) effects and total effects for these exposure-mediator pathways, in each simulation. To do so, we followed the following steps:

\begin{enumerate}
\item We generated a random sample of size $n$, which we divided into $K$ equal-sized estimation samples of size $n_k = n/K$, each with a corresponding parameter generating sample of size $n - n_k$.
\item At each iteration, we used the parameter generating sample to define the mediation pathway(s) and create the estimators for the nuisance parameters used for $\theta(\delta)_d$ and $\theta(\delta)_t$. We then use the estimation sample to obtain the causal parameter estimate using generating equations and TMLE. We repeated this process for all folds.
\item At each iteration, we output the stochastic shift estimates given the pooled one-step and TMLE estimation.
\item For the simple DGP, we use the var\_sets parameter in SuperNOVA to bypass the data-adaptive discovery of mediating paths and simply examine performance of the one $A-Z$ pathway. For the complicated DGP we use the discover\_only parameter to do only pathway discovery and skip estimation. In the complicated DGP we report the proportion of iterations NOVAPathways identifies the correct two pathways out of the possible twenty-five pathways.
\end{enumerate}

To evaluate the performance of our approach, we calculated several metrics for each iteration, including bias, variance, MSE, confidence interval (CI) coverage, and the proportion of instances in which the true meditating pathways were identified. To visually inspect if the rate of convergence was at least as fast as $\sqrt{n}$, we show projections of a $\sqrt{n}$ consistent estimator starting from the initial bias. For brevity, we focus on the absolute bias and confidence interval coverage. We calculated these performance metrics at each iteration, performing 50 iterations for each sample size $n =$ (250, 500, 1000, 1500, 2000, 2500, 3000). We used SuperNOVA with 10-fold cross-validation and default learner stacks for each nuisance parameter and data-adaptive parameter. Additionally, the quantile threshold was set to 0 to include all basis functions used in the final best fitting model. To ensure our estimator has a sampling distribution that is normal, we standardize the bias by dividing by the standard deviation of the estimate at each sample size and plot the density distributions for the direct, indirect and total effects.

\subsection{Default Estimators}

SuperNOVA has two density estimating methods that come built into the package. The haldensify estimator \cite{haldensify} can be used for conditional density estimation of $g_n = p(A|W)$, $e_n = p(A|W,Z)$ and $r_n = p(Z|W)$. Haldensify is a flexible, data-adaptive approach that employs a histogram-based technique to estimate densities. The maximum interaction degree is set by the user as is the number of bins to discretize the outcome. Haldensify works by constructing a histogram of the data and employing a multivariate step function to estimate the density, which makes it computationally efficient and suitable for a wide range of applications.

As an alternative to haldensify, the SuperNOVA package also offers the option to use Super Learner for conditional density estimation. The default Super Learner stack includes a diverse set of learners, such as glm \cite{glm}, elastic net \cite{elasticnet}, random forest \cite{ranger}, and xgboost \cite{xgboost}. We create estimators based on homoscedastic errors (HOSE) and heteroscedastic errors (HESE). For the simulations presented in this paper, we have opted to use Super Learner. This way we can investigate the behavior of the estimator when the true function or an algorithm that approximates the true function is included in the Super Learner library.

Additionally, we need an estimator for $\bar{Q} = E(Y|A,W)$. SuperNOVA provides default algorithms to be used in a Super Learner \cite{SL_2008} that are both fast and flexible. For our data-adaptive procedure, we include learners from the packages earth \cite{earth}, polspline \cite{polspline}, and hal9001 \cite{hal9001}. The results from each of these packages can be formed into a model matrix, on which we can fit an ANOVA to obtain the resulting linear model of basis functions.

In the case where $A$ is discrete, $g_n = p(A|W)$ and $e_n = p(A|W,Z)$ are instead Super Learners built from categorical outcome estimators such as neural networks, random forest and polspline. 

\subsection{Results}

\subsubsection{Do Target Parameters Estimated by NOVAPathways Converge to Truth at $1/\sqrt{n}$ for Continuous Exposures?}

An important aspect of our estimator's performance is its convergence rate. In the context of our simulation (DGP 1 with one exposure-mediator pathway), the convergence rate signifies how quickly the estimator approaches the true parameter value as the sample size increases. Ideally, we want estimates for the total effect, direct effect and indirect effect to show convergence to the truth at $\sqrt{n}$ using a DGP that, although simple, at least includes confounding and relationships that feasible could be observed in a real-world analysis setting. 

\begin{figure}[h]
  \hspace*{-1cm}\includegraphics[scale = 0.5]{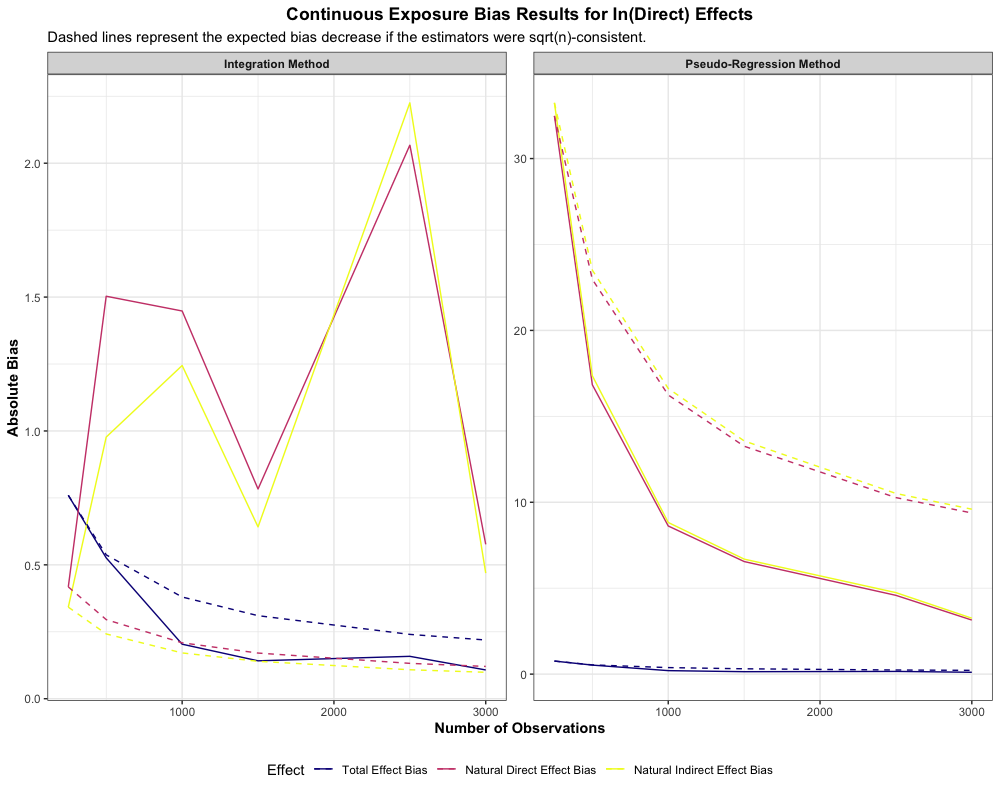}
  \caption{Absolute Bias and Expected $\sqrt{n}$ Convergence Across Sample Sizes for Total, Natural Direct and Natural Indirect Effects when Exposure is Continuous in DGP 1}
  \label{fig:cont_convergence}
\end{figure}

\textbf{Figure \ref{fig:cont_convergence}} exhibits the absolute bias and the anticipated rate of convergence for a $\sqrt{n}$ consistent estimator, given the initial bias, when the exposure is truly continuous. It shows the bias as the sample size increases to 3000. Observing the estimates from the integration method, the bias is generally lower but exhibits a non-convergent behavior when reaching a sample size of 3000, particularly for Natural Direct Effect (NDE) and Natural Indirect Effect (NIE). Although the pseudo-regression approach displays greater consistency, the bias remains considerably high, hindering proper coverage.

\begin{figure}[h]
  \hspace*{-1cm}\includegraphics[scale = 0.5]{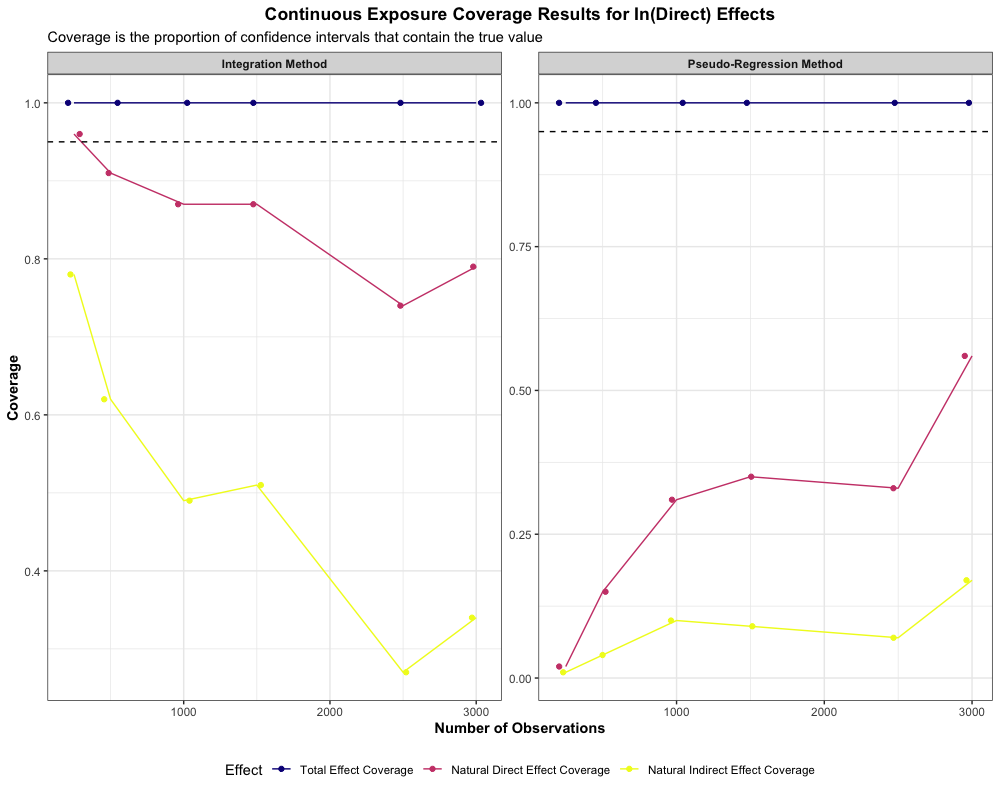}
  \caption{Confidence Interval Coverage for Total, Direct and Indirect Estimates using Integration and Pseudo-Regression in DGP 1}
  \label{fig:cont_CI_coverage}
\end{figure}

Coverage, illustrated in \textbf{Figure \ref{fig:cont_CI_coverage}}, refers to the proportion of iterations for each sample size where confidence intervals contain the true value. For both methodologies—integration and pseudo-regression—the estimated coverage for NDE and NIE does not achieve the desired 95\% level. This shortfall is likely attributable to the bias in estimates induced by numeric integration, a necessary procedure for estimating the nuisance parameters in the case of a continuous exposure. 

Therefore, continuous exposures do not demonstrate the expected $\sqrt{n}$ convergence. This behavior implies that our estimator falls short of the necessary criteria to qualify as asymptotically normal. The departure from asymptotic normality may partly stem from approximations made during the numerical integration required for our estimation process. Alternatively, coding inaccuracies could be at play. Theoretically, the estimator should function correctly, so these anomalies warrant further investigation. Additionally, a sample size of 3000 may still be too small to assess for normality.

\subsubsection{Do Target Parameters Estimated by NOVAPathways Converge to Truth at $1/\sqrt{n}$ for Quantized Exposures?}

We also evaluated the performance of the NOVAPathways estimator under our DGP 1 scenario where the exposure variable is quantized. Like for the truly continuous exposure, the main aspects of the evaluation are the rate of convergence and the coverage of the confidence intervals, as these metrics represent the robustness and reliability of the estimator.

\begin{figure}[!h]
  \hspace*{-1cm}\includegraphics[scale = 0.5]{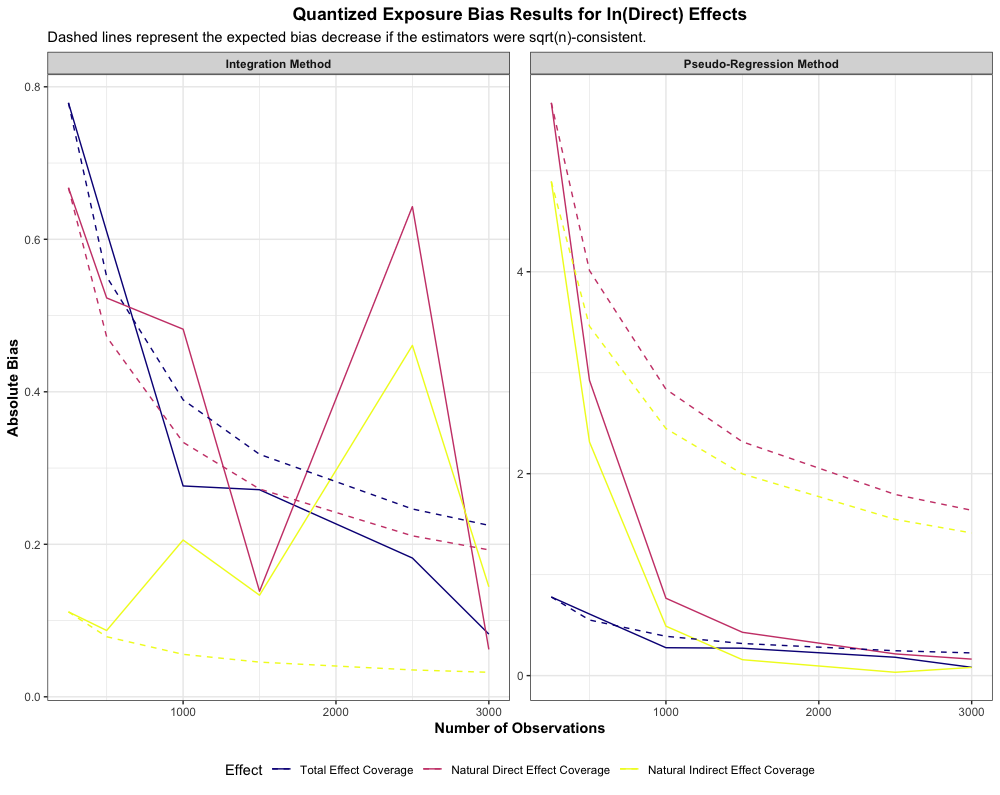}
  \caption{Absolute Bias and Expected $\sqrt{n}$ Convergence Across Sample Sizes for Total, Natural Direct and Natural Indirect Effects when Exposure is Quantized in DGP 1}
  \label{fig:discrete_bias}
\end{figure}

In contrast to the results for continuous exposures, we observe satisfactory performance of the estimator for quantized exposures. The bias for the estimated NDE, NIE, and total effect demonstrates a clear trend of convergence towards zero with increasing sample size, for both integration and pseudo-regression methods. \textbf{Figure \ref{fig:discrete_bias}} shows the absolute bias and expected $\sqrt{n}$ convergence given initial bias as sample size increases. This pattern is more prominent for the integration method, with bias levels generally being much lower than in pseudo-regression. For instance, for NDE, the absolute bias using the integration method decreases from $0.668$ at a sample size of $250$ to $0.0621$ at a sample size of $3000$. 

\begin{figure}[!h]
  \hspace*{-1cm}\includegraphics[scale = 0.5]{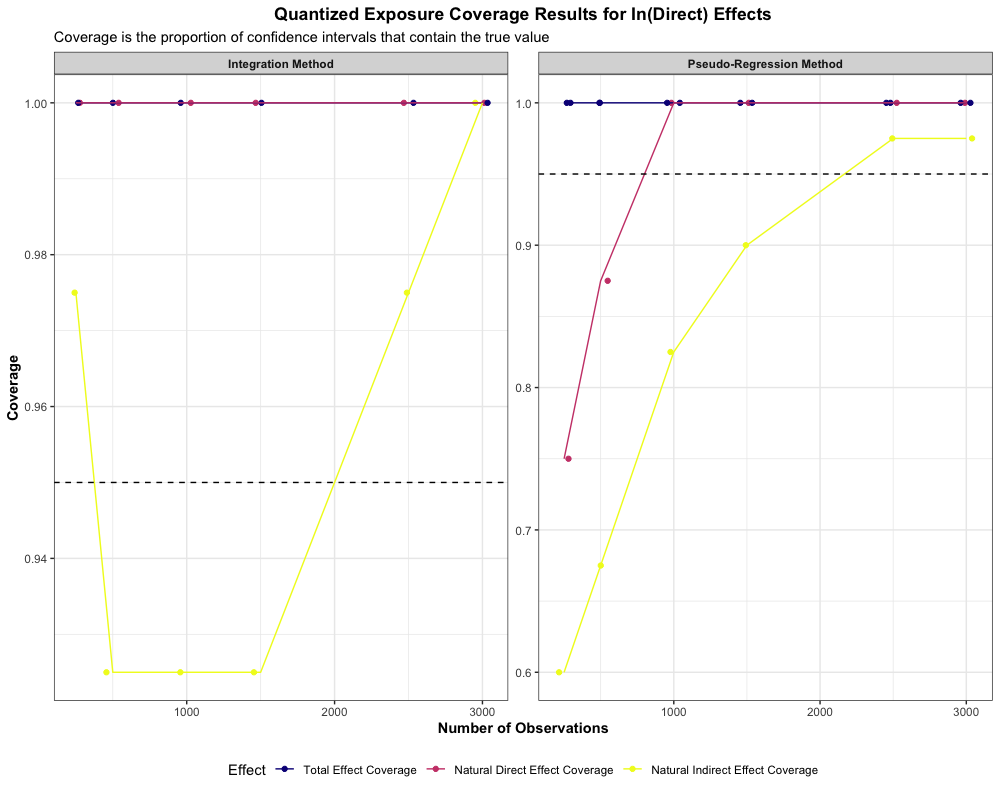}
  \caption{Confidence Interval Coverage for Total, Direct and Indirect Estimates using Integration and Pseudo-Regression in DGP 1}
  \label{fig:discrete_CI_coverage}
\end{figure}

Coverage of confidence intervals for these estimates also shows a desirable pattern. Considering the average coverage across all sample sizes, the coverage for NDE reaches an average of 95.6 \% for the pseudo-regression method and remains at 100\% for the integration method. For NIE, the pseudo-regression method provides an average coverage of 85\%, while the integration method provides a higher average coverage of 96\%. \textbf{Figure \ref{fig:discrete_CI_coverage}} shows the proportion of confidence intervals that contain the true value for each approach at increasing sample size. Finally, for the total effect, the average coverage across all sample sizes reaches 100\%. It's worth noting that for both NDE and NIE, the pseudo-regression method exhibits lower coverage than the integration method.

In summary, our results demonstrate that when the exposure variable is quantized, the NOVAPathways estimator exhibits desirable characteristics of a reliable estimator. It provides a rate of convergence that meets the $1/\sqrt{n}$ standard, and the confidence intervals demonstrate appropriate coverage. These results confirm the robustness of NOVAPathways when applied to quantized exposure variables, and underline the necessity of having appropriately quantized exposure variables in order to achieve reliable and valid results.

\subsubsection{NOVAPathways Correctly Identifies Mediating Pathwways in a Realistic Complex Mixed Exposure-Mediator Situation}

\begin{figure}[!h]
  \hspace*{-2cm}\includegraphics[scale = 0.6]{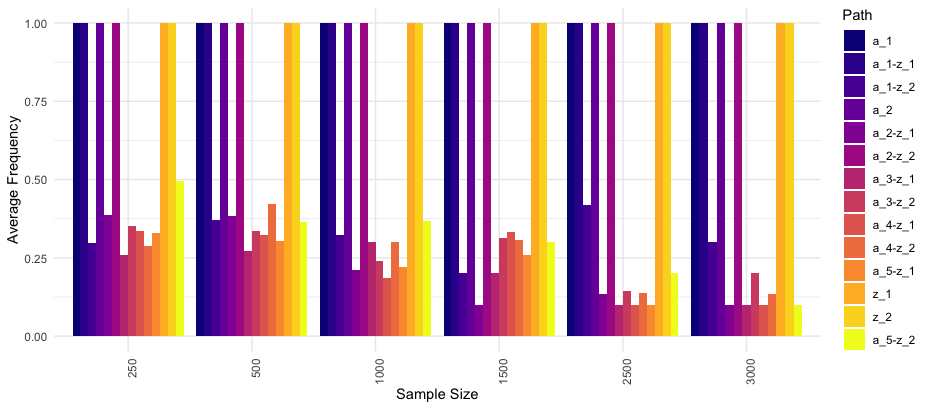}
  \caption{Average Frequency Each Path Was Detected in the Mixed Exposure-Mediator Simulation in DGP 2}
  \label{fig:path_detection}
\end{figure}

In DGP 2, despite having 25 potential mediating pathways in the complex exposure mixture-mediation simulation, NOVAPathways consistently identified the two true pathways ($A_1-Z_1$ and $A_2-Z_2$) with a frequency of 100\%, across various sample sizes ranging from 250 to 3000 observations. Furthermore, direct effects of $A_1$ and $A_2$ on the outcome $Y$ were also consistently identified across all scenarios, reinforcing the robustness of our detection method. \textbf{Figure \ref{fig:path_detection}} shows the frequency each pathways was detected for each sample size. Note that, only pathways detected are reported. 

However, it is noteworthy that there were some instances of incorrectly identified pathways, as seen from the non-zero frequencies of pathways such as $A_1-Z_2$, $A_2-Z_1$, and others, which could be attributed to the high correlation between the exposures. While these false discoveries present opportunities for methodological refinement, the consistently correct identification of the true pathways underpins the effectiveness of our methodology in the presence of multiple mediators and exposures, which is a common scenario in air pollution research. Of note is that, the incorrect pathways were identified in very few folds and in such cases the analyst would report the inconsistency of such a finding. 

\subsubsection{Assessing the Validity of NOVAPathways's Inference through Simulations}

\begin{figure}[!h]
  \hspace*{-2cm}\includegraphics[scale = 0.55]{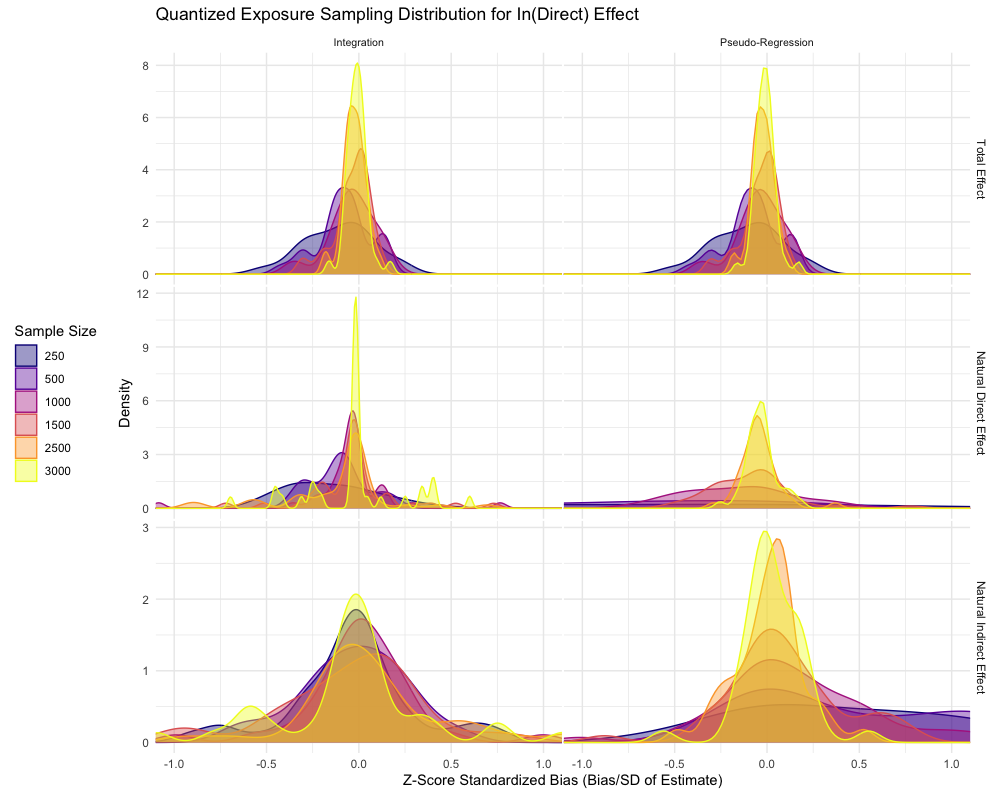}
  \caption{Bias Standardized by Standard Error of Estimates for the Natural In(Direct) Effects and Total Effect in DGP 1}
  \label{fig:z_dist}
\end{figure}

The fundamental premise of a robust inference is the verification of the estimator's normal sampling distribution, centered at zero and progressively narrowing with increasing sample size. This premise is tested in the context of the NOVAPathway estimator for natural direct, indirect, and total effects. We illustrate the empirical distribution of the standardized bias, defined as the difference between the estimated and true values from the data-generating process, normalized by the standard deviation of the estimates across iterations. The assessment is conducted using 50 iterations per sample size and visualized as a probability density distribution in \textbf{Figure \ref{fig:z_dist}}.

In \textbf{Figure \ref{fig:z_dist}}, we observe the convergence of the sampling distribution to a mean-zero normal as sample size escalates. This phenomenon is evident across all types of effect estimates. The total effect, calculated via one-step, remains consistent regardless of whether the natural direct effect (NDE) is computed using integration or pseudo-regression methods. The NDE for the integration method is concentrated more closely around zero, albeit exhibiting greater tail variability, while the pseudo-regression counterpart maintains a smoother, centered distribution. The natural indirect effect (NIE) demonstrates the widest dispersion, although still centered around zero. Notably, the pseudo-regression method achieves a slightly narrower distribution around zero for NIE.

All plots in \textbf{Figure \ref{fig:z_dist}} exhibit normal or near-normal distributions centered at zero that contract with an increase in sample size. This characteristic is crucial for the validity of confidence interval construction and underscores the reliability of our estimator. As such, the simulation results affirm the soundness of NOVAPathways's inference methodology.

\section{Applications}

\subsection{NHANES Data}

\subsubsection{Data Description}
To provide a motivating example for the application of NOVAPathways we extracted data from the 2001-2002 cycle of the National Health and Nutrition Examination Survey (NHANES). The NHANES program, managed by the Centers for Disease Control and Prevention (CDC), is a comprehensive set of studies designed to assess the health and nutritional status of adults and children in the United States \cite{zipf2013national}. These studies employ a combination of interviews and physical examinations to capture a broad array of health information. NHANES data is particularly suitable for motivating the use of NOVAPathways due to its representative sample of the U.S. population (specifically for pollution exposure), broad collection of health-related variables, and its open availability. This enables us to make our analysis transparent and easily replicable, fostering open science practices and facilitating methodological testing \cite{wilson2017good}. For these purposes, all code for data cleaning and curation for this motivating analysis example using NHANES is included in the SuperNOVA package which uses the NOVAPathways method. 

One significant challenge of using cross-sectional datasets like NHANES is the potential for reverse causality, wherein the outcomes/mediators may influence the exposures rather than vice versa. This characteristic violates the temporal assumption required for traditional causal inference \cite{hernan2010causal}. However, the use of NHANES data in our study is not primarily to establish causal relationships but rather to provide a real-world demonstration of the capabilities of our method, NOVAPathways. 

The NHANES data provides a large number of well measured toxic metal exposures, biomarkers for possible mediating pathways and covariates. For our purposes, this offers an opportunity to determine if NOVAPathways identifies consistent mediating pathways in high-dimensional data and delivers interpretable direct, indirect and total effect results based on stochastic shift interventions data-adaptively determined pathways. For our motivating example we investigate the association of a mixture of toxic metals on asthma both directly and possibly indirectly through biomarkers for inflammation, oxidative stress and immune function. 

Our choice of the 2001-2002 NHANES data cycle was informed by the fact that this cycle included all relevant variables necessary for a comprehensive investigation into the associations between toxic metal exposures, inflammation, immune function, oxidative stress, and the prevalence of asthma \cite{Balali_Mood2021}. This particular NHANES cycle collected exhaustive data on these variables, offering a unique opportunity to conduct our investigation within a representative sample of the U.S. population.

The original NHANES 2001-2002 dataset consisted of 11,039 participants, with 4,260 individuals providing blood samples and consenting to DNA analysis \cite{zipf2013national}. After applying our exclusion criteria, such as missing environmental chemical analysis data, missing key covariate data, and insufficient stored samples for telomere length estimation, our final study sample was comprised of 1,344 participants.

Our data cleaning and curation techniques were relatively basic as our main goal is to demonstrate our proposed methodology and software, not provide a thorough analysis. Nontheless, data cleaning and curation were undertaken to ensure the integrity of distributions in our dataset was retained while allowing us to not lose too many observations due to missingness. We first omitted observations with missing values in the outcome variable (asthma) and in the crucial exposure variables (toxic metals). We then retained columns where less than 20\% of the data was missing. This balance allowed us to maximize the use of available data while avoiding the potential bias from imputing excessive missing values.

Next, we imputed missing data in the remaining variables through suitable methods: mean imputation for numeric variables and mode imputation for categorical ones \cite{horton2007much}. This strategy helped ensure the final dataset maintained the original distributions and variable relationships to the greatest extent possible.

We then quantized the metal exposure data to address the methodological issue with our proposed method when the exposure is fully continuous. As shown, continuous exposures necessitate numeric integration in the calculations of the mediation effects. However, this approach lead to approximations that are not precise enough, inducing asymptotic bias and resulting in poor confidence interval coverage. To avoid this issue, we quantized the continuous exposure data, transforming each exposure into a categorical variable with equal frequency bins (deciles in our case). This transformation allows a shift delta = 1 to represent an increase in decile, and we can calculate each nuisance function as a simple weighted sum rather than a numeric integration. By doing this, we have observed improved asymptotic behavior of our estimators and accurate confidence interval coverage.

The selection of toxic metals as exposures in our study was informed by prior literature demonstrating the potential link between toxic metal exposure, oxidative stress, inflammation, and immune function—all factors implicated in the etiology of asthma \cite{grzela2017oxidative, vargas2021metals}. Several studies have shown that exposure to toxic metals can lead to oxidative stress, which in turn can trigger inflammatory responses and modulate immune function \cite{esmaeilzadeh2021role, Kong2023, Rahman2006}. These processes can potentially contribute to the onset or exacerbation of asthma, hence our interest in exploring these relationships in this study. By investigating these associations within the NHANES 2001-2002 dataset, we aim to show that semi-parametric methods utilizing efficient estimators and data-adaptive target parameters can yield a deeper understanding of the complex interplay between environmental exposures, molecular biomarkers, and disease outcomes.

Through this process, we seek to illustrate the utility of our SuperNOVA software which incorporates the NOVAPathways mediation methodology. Our results, which are presented in subsequent tables, offer a comprehensive view of the information that SuperNOVA can generate from a provided dataset.

\subsubsection{Consistent Findings for Toxic Metal Exposure on Asthma Through Inflammatory, Oxidative Stress and Immune Function Mediators}

Because NOVAPathways data-adaptively discovers exposure-mediator pathways, it's best to report first any notable consistencies across the multiple folds. Cesium, as an exposure, was found in 80\% of the folds, demonstrating the greatest consistency among all exposures investigated. This highlights the potential relevance of cesium in our model, warranting further investigations into its role and impact on asthma in future studies. 

When considering the mediators, monocyte percentage and vitamin E emerged as the most consistent across the folds, being detected in 80\% and 60\% of the folds, respectively. The consistent appearance of monocyte percentage, a key indicator of immune system activation, underscores the possible involvement of immune modulation in the effect of toxic metal exposure on asthma. Similarly, the recurring detection of vitamin E may imply a role for antioxidant mechanisms in modulating the exposure-asthma relationship.

Furthermore, we found the exposure-mediator pairs of cesium-monocyte percentage and tungsten-monocyte percentage in 60\% of the folds. The pairings of specific exposures with monocyte percentage suggest potential pathways where these elements could influence asthma pathogenesis through immune mechanisms. Meanwhile, the lead-vitamin E pair appeared in 50\% of the folds, alluding to another potential pathway via oxidative stress mechanisms.

Given these results, we next report the fold specific and pooled results of Cesium, the Cesium-Monocyte Percentage pathway, the Lead-Vitamin E pathway, and the Tungsten-Monocyte Percentage pathway. 

\subsubsection{Results for Cesium, Lead and Tungsten Through Monocyte Percentage and Vitamin E}

We first examined the potential impact of Cesium exposure on the likelihood of developing asthma, independent of mediation. In 8 out of 10 folds, Cesium consistently appeared, implying a possible influence in the disease's progression. Here we use a decile shift increment in Cesium exposure and observe the expected probability of asthma given this shift compared to the observed probability of asthma. Here, a decile increase is equivalent to a rise of $4.182,\upmu$/L on the Cesium continuous scale.

While the results varied slightly across the folds, the effect generally leaned towards the positive. In the pooled analysis, a decile increase in Cesium corresponded to a 0.012 increase in the asthma probability. However, this result didn't achieve statistical significance at the conventional 0.05 level (p-value = 0.17). While the findings do not conclusively establish a relationship between cesium and asthma, the consistent results hint at a potential correlation warranting further investigation. \textbf{Table \ref{tab:cesium_total_effect}} presents the total effects of a decile shift in Cesium on the likelihood of asthma.

\begin{table}[h]
\centering
\hspace*{-2cm}\begin{tabular}{rlrrrrrrlllrr}
  \toprule
  Psi & Variance & SE & Lower CI & Upper CI & P-value & Fold & Type & Variables & N & Delta \\ 
  \midrule
  0.02 & 0.00 & 0.01 & -0.01 & 0.04 & 0.14 & 1 & Indiv Shift & cesium & 135.00 & 1.00 \\ 
  0.02 & 0.00 & 0.05 & -0.07 & 0.12 & 0.64 & 4 & Indiv Shift & cesium & 135.00 & 1.00 \\ 
  0.04 & 0.00 & 0.03 & -0.03 & 0.10 & 0.29 & 5 & Indiv Shift & cesium & 135.00 & 1.00 \\ 
  -0.00 & 0.00 & 0.02 & -0.04 & 0.04 & 0.92 & 6 & Indiv Shift & cesium & 134.00 & 1.00 \\ 
  0.00 & 0.00 & 0.01 & -0.02 & 0.02 & 0.85 & 7 & Indiv Shift & cesium & 134.00 & 1.00 \\ 
  0.00 & 0.00 & 0.01 & -0.01 & 0.01 & 0.88 & 8 & Indiv Shift & cesium & 134.00 & 1.00 \\ 
  0.02 & 0.00 & 0.02 & -0.03 & 0.06 & 0.42 & 9 & Indiv Shift & cesium & 134.00 & 1.00 \\ 
  0.00 & 0.00 & 0.01 & -0.01 & 0.02 & 0.74 & 10 & Indiv Shift & cesium & 133.00 & 1.00 \\ 
  0.01 & 0.00 & 0.01 & -0.01 & 0.03 & 0.17 & Pooled TMLE & Indiv Shift & cesium & 1074.00 & 1.00 \\ 
   \bottomrule
\end{tabular}
\caption{Results for Association Between a Decile Shift in Cesium and Probability of Asthma}
\label{tab:cesium_total_effect}
\end{table}

We subsequently examined how much of this effect passed through the monocyte percentage as opposed to not. \textbf{Table \ref{tab:cesium_monocyte}} displays the fold-specific and pooled results for the NDE, NIE, and total effect of Cesium on asthma, via monocyte percentage, using both pseudo-regression and double integration methods to construct the estimator. With pseudo-regression estimates, we observed an NDE of 0.61 (-0.66 - 1.88) and an NIE through monocyte percentage of -0.60 (-1.86 - 0.66). Both results were not significant. However, despite the absence of traditional statistical significance, the persistence of cesium through the monocyte percentage across the majority of the folds suggests a potential influence of this pathway on asthma.

\begin{table}[ht]
\centering
\begin{tabular}{rllrrrrrr}
  \toprule
 Fold & Parameter & Psi & Variance & SE & Lower CI & Upper CI & P-Value \\ 
  \midrule
  Fold 1  & NDE-Pseudo-Reg & 0.05 & 0.00 & 0.06 & -0.07 & 0.17 & 0.38 \\ 
  Fold 1  & NDE-Double-Int & 0.05 & 0.00 & 0.06 & -0.07 & 0.17 & 0.41 \\ 
  Fold 1  & NIE-Pseudo-Reg & -0.04 & 0.00 & 0.06 & -0.16 & 0.07 & 0.46 \\ 
  Fold 1  & NIE-Double-Int & -0.04 & 0.00 & 0.06 & -0.16 & 0.08 & 0.50 \\ 
  Fold 1  & Total Effect & 0.01 & 0.00 & 0.01 & -0.01 & 0.03 & 0.41 \\ \hline
  Fold 4  & NDE-Pseudo-Reg & 2.66 & 7.64 & 2.76 & -2.76 & 8.08 & 0.34 \\ 
  Fold 4  & NDE-Double-Int & 2.66 & 7.64 & 2.76 & -2.76 & 8.07 & 0.34 \\ 
  Fold 4  & NIE-Pseudo-Reg & -2.62 & 7.66 & 2.77 & -8.05 & 2.80 & 0.34 \\ 
  Fold 4  & NIE-Double-Int & -2.62 & 7.66 & 2.77 & -8.04 & 2.80 & 0.34 \\ 
  Fold 4  & Total Effect & 0.04 & 0.00 & 0.05 & -0.06 & 0.13 & 0.45 \\ \hline
  Fold 5  & NDE-Pseudo-Reg & 0.92 & 7.34 & 2.71 & -4.39 & 6.23 & 0.73 \\ 
  Fold 5  & NDE-Double-Int & 0.92 & 7.34 & 2.71 & -4.39 & 6.22 & 0.74 \\ 
  Fold 5  & NIE-Pseudo-Reg & -0.88 & 7.16 & 2.67 & -6.12 & 4.36 & 0.74 \\ 
  Fold 5  & NIE-Double-Int & -0.87 & 7.16 & 2.67 & -6.12 & 4.37 & 0.74 \\ 
  Fold 5  & Total Effect & 0.04 & 0.00 & 0.04 & -0.03 & 0.11 & 0.28 \\ \hline
  Fold 6  & NDE-Pseudo-Reg & -0.02 & 0.00 & 0.04 & -0.10 & 0.07 & 0.72 \\ 
  Fold 6  & NDE-Double-Int & -0.02 & 0.00 & 0.04 & -0.11 & 0.07 & 0.64 \\ 
  Fold 6  & NIE-Pseudo-Reg & -0.02 & 0.00 & 0.06 & -0.14 & 0.10 & 0.73 \\ 
  Fold 6  & NIE-Double-Int & -0.01 & 0.00 & 0.06 & -0.13 & 0.10 & 0.80 \\ 
  Fold 6  & Total Effect & -0.04 & 0.00 & 0.03 & -0.09 & 0.02 & 0.18 \\ \hline
  Fold 9  & NDE-Pseudo-Reg & 0.03 & 0.00 & 0.02 & -0.02 & 0.07 & 0.26 \\ 
  Fold 9  & NDE-Double-Int & 0.02 & 0.00 & 0.02 & -0.02 & 0.07 & 0.30 \\ 
  Fold 9  & NIE-Pseudo-Reg & -0.02 & 0.00 & 0.03 & -0.07 & 0.04 & 0.57 \\ 
  Fold 9  & NIE-Double-Int & -0.01 & 0.00 & 0.03 & -0.07 & 0.04 & 0.62 \\ 
  Fold 9  & Total Effect & 0.01 & 0.00 & 0.02 & -0.03 & 0.04 & 0.64 \\ \hline
  Fold 10  & NDE-Pseudo-Reg & -0.01 & 0.00 & 0.01 & -0.04 & 0.02 & 0.67 \\ 
  Fold 10  & NDE-Double-Int & -0.01 & 0.00 & 0.01 & -0.04 & 0.02 & 0.47 \\ 
  Fold 10  & NIE-Pseudo-Reg & 0.01 & 0.00 & 0.02 & -0.04 & 0.05 & 0.72 \\ 
  Fold 10  & NIE-Double-Int & 0.01 & 0.00 & 0.02 & -0.03 & 0.06 & 0.59 \\ 
  Fold 10  & Total Effect & 0.00 & 0.00 & 0.01 & -0.03 & 0.03 & 0.89 \\ \hline
  Pooled & NDE-Pseudo-Reg & 0.61 & 0.42 & 0.65 & -0.66 & 1.88 & 0.35 \\ 
  Pooled & NDE-Integrated & 0.60 & 0.42 & 0.65 & -0.66 & 1.87 & 0.35 \\ 
  Pooled & NIE-Pseudo-Reg & -0.60 & 0.41 & 0.64 & -1.86 & 0.66 & 0.35 \\ 
  Pooled & NIE-Integrated & -0.59 & 0.41 & 0.64 & -1.86 & 0.67 & 0.35 \\ 
  Pooled & Total-Pooled-TMLE & 0.01 & 0.00 & 0.01 & -0.01 & 0.03 & 0.40 \\ 
   \bottomrule
\end{tabular}
\caption{NDE and NIE of Cesium on Asthma Through Monocyte Percentage Across the Folds}
\label{tab:cesium_monocyte}
\end{table}

For Lead and Tungsten, a similar process was followed. In this instance, a decile increase in Lead and Tungsten corresponded to a rise of $1.3,\upmu$/L and $0.285,\upmu$/L on the respective continuous scales. A pathway for Tungsten through monocyte percentage was found in 60\%. These results are provided in \textbf{Table \ref{tab:tungston_monocyte}}. Like Cesium-monocyte percentage, although these pathways were found in a majority of folds, the effects were not significant. A one decile increase in Tungston is associated with a -0.005 (-0.013 - 0.004) decrease in the probability of asthma (p-value = 0.28). Results for the NDE using pseudo-regression are the same as the total effect indicating no indirect effect through monocyte percentage. 

\begin{table}[ht]
\centering
\begin{tabular}{rllrrrrrr}
  \toprule
 Fold & Parameter & Psi & Variance & SE & Lower CI & Upper CI & P-Value \\ 
  \midrule
  Fold 4  & NDE-Pseudo-Reg & 0.01 & 0.00 & 0.02 & -0.02 & 0.05 & 0.40 \\ 
  Fold 4  & NDE-Double-Int & 0.01 & 0.00 & 0.02 & -0.02 & 0.05 & 0.37 \\ 
  Fold 4  & NIE-Pseudo-Reg & -0.02 & 0.00 & 0.02 & -0.06 & 0.02 & 0.35 \\ 
  Fold 4  & NIE-Double-Int & -0.02 & 0.00 & 0.02 & -0.06 & 0.02 & 0.33 \\ 
  Fold 4  & Total Effect & -0.01 & 0.00 & 0.01 & -0.03 & 0.01 & 0.57 \\ \hline
  Fold 5  & NDE-Pseudo-Reg & -0.01 & 0.00 & 0.01 & -0.02 & 0.01 & 0.58 \\ 
  Fold 5  & NDE-Double-Int & -0.00 & 0.00 & 0.01 & -0.02 & 0.01 & 0.59 \\ 
  Fold 5  & NIE-Pseudo-Reg & -0.01 & 0.00 & 0.01 & -0.04 & 0.01 & 0.33 \\ 
  Fold 5  & NIE-Double-Int & -0.01 & 0.00 & 0.01 & -0.04 & 0.01 & 0.32 \\ 
  Fold 5  & Total Effect & -0.02 & 0.00 & 0.01 & -0.03 & 0.00 & 0.06 \\ \hline
  Fold 6  & NDE-Pseudo-Reg & -0.01 & 0.00 & 0.01 & -0.03 & 0.00 & 0.14 \\ 
  Fold 6  & NDE-Double-Int & -0.01 & 0.00 & 0.01 & -0.03 & 0.00 & 0.15 \\ 
  Fold 6  & NIE-Pseudo-Reg & 0.01 & 0.00 & 0.01 & -0.02 & 0.04 & 0.52 \\ 
  Fold 6  & NIE-Double-Int & 0.01 & 0.00 & 0.01 & -0.02 & 0.04 & 0.54 \\ 
  Fold 6  & Total Effect & -0.00 & 0.00 & 0.01 & -0.02 & 0.02 & 0.69 \\ \hline
  Fold 7  & NDE-Pseudo-Reg & -0.01 & 0.00 & 0.01 & -0.03 & -0.00 & 0.03 \\ 
  Fold 7  & NDE-Double-Int & -0.01 & 0.00 & 0.01 & -0.03 & -0.00 & 0.04 \\ 
  Fold 7  & NIE-Pseudo-Reg & 0.01 & 0.00 & 0.01 & -0.02 & 0.04 & 0.46 \\ 
  Fold 7  & NIE-Double-Int & 0.01 & 0.00 & 0.01 & -0.02 & 0.04 & 0.50 \\ 
  Fold 7  & Total Effect & -0.00 & 0.00 & 0.01 & -0.03 & 0.02 & 0.72 \\ \hline
  Fold 9  & NDE-Pseudo-Reg & 0.01 & 0.00 & 0.01 & -0.01 & 0.03 & 0.48 \\ 
  Fold 9  & NDE-Double-Int & 0.01 & 0.00 & 0.01 & -0.02 & 0.03 & 0.56 \\ 
  Fold 9  & NIE-Pseudo-Reg & -0.00 & 0.00 & 0.01 & -0.01 & 0.01 & 0.71 \\ 
  Fold 9  & NIE-Double-Int & -0.00 & 0.00 & 0.01 & -0.01 & 0.01 & 0.90 \\ 
  Fold 9  & Total Effect & 0.01 & 0.00 & 0.01 & -0.02 & 0.03 & 0.63 \\ \hline
  Fold 10  & NDE-Pseudo-Reg & -0.02 & 0.00 & 0.01 & -0.03 & -0.00 & 0.03 \\ 
  Fold 10  & NDE-Double-Int & -0.02 & 0.00 & 0.01 & -0.03 & -0.00 & 0.03 \\ 
  Fold 10  & NIE-Pseudo-Reg & 0.01 & 0.00 & 0.01 & -0.01 & 0.04 & 0.24 \\ 
  Fold 10  & NIE-Double-Int & 0.01 & 0.00 & 0.01 & -0.01 & 0.04 & 0.25 \\ 
  Fold 10  & Total Effect & -0.00 & 0.00 & 0.01 & -0.02 & 0.01 & 0.64 \\ \hline
  Pooled & NDE-Pseudo-Reg & -0.01 & 0.00 & 0.00 & -0.01 & 0.00 & 0.30 \\ 
  Pooled & NDE-Integrated & -0.00 & 0.00 & 0.00 & -0.01 & 0.00 & 0.32 \\ 
  Pooled & NIE-Pseudo-Reg & 0.00 & 0.00 & 0.01 & -0.01 & 0.01 & 1.00 \\ 
  Pooled & NIE-Integrated & -0.00 & 0.00 & 0.01 & -0.01 & 0.01 & 0.97 \\ 
  Pooled & Total-Pooled-TMLE & -0.01 & 0.00 & 0.00 & -0.01 & 0.00 & 0.28 \\ 
   \hline
\end{tabular}
\caption{NDE and NIE of Tungston on Asthma Through Monocyte Percentage Across the Folds}
\label{tab:tungston_monocyte}
\end{table}

Lastly, we give results for Lead on asthma through vitamin E.  \textbf{Table \ref{tab:lead_vite}} shows these results. Again, the total effect, NDE and NIE are not significant given a decile increase in lead although this pathway was identified in 50\% of the folds. 

\begin{table}[ht]
\centering
\begin{tabular}{rllrrrrrr}
  \toprule
 Fold & Parameter & Psi & Variance & SE & Lower CI & Upper CI & P-Value \\ 
  \midrule
  Fold 1  & NDE-Pseudo-Reg & -0.02 & 0.00 & 0.03 & -0.08 & 0.03 & 0.43 \\ 
  Fold 1  & NDE-Double-Int & -0.02 & 0.00 & 0.03 & -0.08 & 0.03 & 0.41 \\ 
  Fold 1  & NIE-Pseudo-Reg & 0.01 & 0.00 & 0.03 & -0.06 & 0.08 & 0.72 \\ 
  Fold 1  & NIE-Double-Int & 0.01 & 0.00 & 0.03 & -0.05 & 0.08 & 0.69 \\ 
  Fold 1  & Total Effect & -0.01 & 0.00 & 0.01 & -0.04 & 0.02 & 0.48 \\ \hline
  Fold 2  & NDE-Pseudo-Reg & -0.02 & 0.00 & 0.02 & -0.06 & 0.01 & 0.17 \\ 
  Fold 2  & NDE-Double-Int & 0.03 & 0.00 & 0.03 & -0.03 & 0.09 & 0.35 \\ 
  Fold 2  & NIE-Pseudo-Reg & 0.03 & 0.00 & 0.03 & -0.03 & 0.08 & 0.33 \\ 
  Fold 2  & NIE-Double-Int & -0.03 & 0.00 & 0.04 & -0.10 & 0.05 & 0.47 \\ 
  Fold 2  & Total Effect & 0.00 & 0.00 & 0.01 & -0.02 & 0.03 & 0.84 \\ \hline
  Fold 3  & NDE-Pseudo-Reg & 0.03 & 0.01 & 0.08 & -0.12 & 0.18 & 0.72 \\ 
  Fold 3  & NDE-Double-Int & 0.03 & 0.01 & 0.08 & -0.13 & 0.18 & 0.74 \\ 
  Fold 3  & NIE-Pseudo-Reg & -0.02 & 0.01 & 0.07 & -0.17 & 0.12 & 0.77 \\ 
  Fold 3  & NIE-Double-Int & -0.02 & 0.01 & 0.07 & -0.16 & 0.12 & 0.78 \\ 
  Fold 3  & Total Effect & 0.01 & 0.00 & 0.01 & -0.02 & 0.03 & 0.61 \\ \hline
  Fold 6  & NDE-Pseudo-Reg & -0.01 & 0.00 & 0.01 & -0.03 & 0.02 & 0.61 \\ 
  Fold 6  & NDE-Double-Int & -0.01 & 0.00 & 0.01 & -0.03 & 0.01 & 0.52 \\ 
  Fold 6  & NIE-Pseudo-Reg & -0.01 & 0.00 & 0.04 & -0.07 & 0.07 & 0.89 \\ 
  Fold 6  & NIE-Double-Int & -0.00 & 0.00 & 0.04 & -0.07 & 0.07 & 0.93 \\ 
  Fold 6  & Total Effect & -0.01 & 0.00 & 0.03 & -0.07 & 0.04 & 0.70 \\ \hline
  Fold 9  & NDE-Pseudo-Reg & -0.06 & 0.00 & 0.05 & -0.16 & 0.03 & 0.19 \\ 
  Fold 9  & NDE-Double-Int & -0.06 & 0.00 & 0.05 & -0.16 & 0.03 & 0.18 \\ 
  Fold 9  & NIE-Pseudo-Reg & 0.04 & 0.00 & 0.06 & -0.07 & 0.15 & 0.44 \\ 
  Fold 9  & NIE-Double-Int & 0.04 & 0.00 & 0.06 & -0.07 & 0.15 & 0.43 \\ 
  Fold 9  & Total Effect & -0.02 & 0.00 & 0.02 & -0.05 & 0.01 & 0.21 \\ \hline
  Pooled & NDE-Pseudo-Reg & -0.02 & 0.00 & 0.02 & -0.06 & 0.02 & 0.37 \\ 
  Pooled & NDE-Integrated & -0.01 & 0.00 & 0.02 & -0.05 & 0.03 & 0.70 \\ 
  Pooled & NIE-Pseudo-Reg & 0.01 & 0.00 & 0.02 & -0.03 & 0.05 & 0.61 \\ 
  Pooled & NIE-Integrated & 0.00 & 0.00 & 0.02 & -0.04 & 0.04 & 0.95 \\ 
  Pooled & Total-Pooled-TMLE & -0.01 & 0.00 & 0.01 & -0.02 & 0.01 & 0.41 \\ 
   \bottomrule
\end{tabular}
\caption{NDE and NIE of Lead on Asthma Through Vitamin E Across the Folds}
\label{tab:lead_vite}
\end{table}

Through an analysis of the NHANES dataset, we highlight the proficiency of NOVAPathways in identifying mediating pathways within high-dimensional data contexts. The dataset in focus included 9 exposures and 12 mediators, thus theoretically encompassing 108 potential pathways. These pathways could mediate the effects of toxic metals via proxies of inflammation, oxidative stress, and immune function.

Using flexible basis estimators, NOVAPathways successfully discerned the most influential pathways and provided estimates associated with a one-decile increment in exposure. While none of the effects reached the threshold of statistical significance, some exhibited borderline significance. 

Our comparison of Natural Direct Effects (NDE) estimates from pseudo-regression and integration procedures revealed similar trends. The stability of estimates across the folds and a decreased variance for the pooled results, as anticipated, reaffirmed the advantage of pooling estimates across folds for precision.

We acknowledge the potential limitations of our method, as we discretized exposures prior to implementing NOVAPathways. This more rudimentary representation of exposures, although simplifying the data, might make pathway discovery more challenging. 

Nevertheless, our primary objective was not the pinpoint accuracy of a causal inference but rather a demonstration of the potential output from NOVAPathways. We sought consistency of results across folds and the provision of interpretable estimates for NDE, NIE, and total effects. In conclusion, this example underscores NOVAPathways' utility in navigating complex associations within high-dimensional data, offering a useful tool for analysts working with multiple exposures and potential mediators.

\section{Software}
The accessibility and application of statistical software that executes semi-parametric methods which respect data-generating processes found in real-world data is pivotal for ensuring consistent and reproducible outcomes across research studies. SuperNOVA, an open-source R package, attempts to address this need by facilitating the evaluation of causal effects from mixed exposures using asymptotically linear estimators, which now includes the NOVAPathways method for mediation. These estimators are proven to converge to the true estimand at $\sqrt{n}$ given estimates of nuisance parameters converge at $n^{1/4}$. Its ability to handle both continuous and discretized exposures addresses a notable limitation of its predecessor, the medshift package \cite{hejazi2020medshift} developed by Ivan Diaz and Nima Hejazi, which only supports binary exposures. We also offer some additional functionality compared to the longitudinal modified treatment policies approach and packege \cite{lmtp_article, lmtp_manual} by data-adaptively finding mediating pathways in cross-sectional data. While continuous exposures are accommodated in SuperNOVA, caution is warranted due to the potential for bias introduced by numerical integration, which we have shown.

At the heart of SuperNOVA, with its integrated NOVAPathways, is Super Learning, a machine learning technique employed via the SL3 package \cite{coyle2018}. This methodology allows SuperNOVA to adaptively identify mediating pathways using ensembles of basis-function estimators, improving the adaptability and efficiency of the software to find pathways even in complex exposure settings. Likewise, Super Learning is used for the estimation of each nuisance function. 

Comparison with existing software illustrates the potential for SuperNOVA to enhance the accuracy and flexibility of mixed exposure-mediator research. Many environmental health studies that have performed mediation analyses have used packages such as medflex \cite{Steen2017} and mediation \cite{mediation_package} which are largely reliant on parametric assumptions. For instance, these packages make strong assumptions about functional form, and they often assume no interactions between the exposure and mediator, which can lead to biased estimates of direct and indirect effects. In contrast, SuperNOVA’s semi-parametric approach relaxes these assumptions, potentially resulting in more accurate and consistent estimates. Additionally, no method or package currently exists which can identify pathways and make valid inference on these pathways in the presence of high-dimensional data. 

SuperNOVA's design allows for both sequential and parallel computing, leveraging the parallel processing capabilities offered by the furrr package \cite{furrr_2022}. Its computational efficiency expands its suitability for use on personal computers, which can be crucial in resource-limited research settings. Additionally, in the context where the analyst has a pre-defined pathways they want to test, the path discovery section of NOVAPathways can be skipped and the direct, indirect and total effects can directly be estimated using the cross-validation procedure. Conversely, if the analyst is instead interested in only finding the most relevant exposure-mediator paths to guide future study develop, this approach is also available.

Additional features of SuperNOVA include a comprehensive vignette, a detailed exposition of the underlying semi-parametric theory, and comparisons to existing methods. The package also offers the NHANES mixed metal exposure data for reproducibility purposes, coding notebooks illustrating the application of the software, and interpretative summaries of SuperNOVA output. SuperNOVA is regularly updated, available on GitHub (https://github.com/blind-contours/SuperNOVA), and aims to equip researchers with robust tools to advance the quality of research in mixed exposure and environmental health.

\section{Limitations}

Even as we have made a concerted effort to apply rigorous methodology in this study, following \cite{diaz_2020} there are several limitations to consider which influenced our results, particularly when the exposure is truly continuous.

Firstly, we used Monte Carlo integration methods, which are inherently stochastic. This could have introduced some level of bias in our estimates. We sought to minimize this by implementing four times the sample size for the number of Monte Carlo samples. However, in high-dimensional or complex model scenarios, such adjustments may not fully eradicate the error.

Furthermore, data variability, particularly in the density estimation, could have contributed to bias introduction. Specifically, when density values hover at the extremes - either exceedingly low or high - the subsequent variance in the estimator may inflate the bias.

Our proposed mediation method for continuous exposures also struggled with potential issues regarding integration boundaries. Even though we were cautious in setting these boundaries (the range of the exposure), the region of integration might have covered areas where the functions integrated were not well-behaved. This could have added to the bias.

Moreover, instabilities in the numerical computations could have subtly influenced our findings. Despite the  power of contemporary computational tools, they are not entirely devoid of errors. Instances of round-off or truncation errors could subtly impact the results.

Likely, this issue with continuous exposures arises as a cumulative effect of the aforementioned limitations. Nevertheless, our results have demonstrated that when exposure is quantized into a discrete form, thereby bypassing numeric integration, our estimator exhibits the expected asymptotic behavior. Moreover, it provides valid confidence intervals for inference - results that can be interpreted continuously.

In relation to positivity, violations of this principle are often an unavoidable reality in many contexts. Nonetheless, our suggested approach optimizes the situation by considering smaller shifts. These shifts are based on the ratio of exposure densities, which contrast the density under shift to the observed density when there is no shift. Similar to our methodology for path discovery, this strategy is heuristic in nature. It attempts to strike a balance between ease of comprehension and implementation while effectively achieving the intended objective.

\section{Discussion}

In this study, we introduce a novel approach for the estimation of natural direct, indirect, and total effects, facilitated through data-adaptive identification of mediating pathways in high-dimensional data. This breakthrough addresses a significant gap in current analytical methods, particularly when dealing with data that comprises numerous exposures and mediators, which is a common occurrence in environmental omics data.

Our approach first fits a very large statistical model to the exposure-mediator-covariate space and treats the basis functions used in this model as a data-adaptive target parameter. This is done in two stages to discover the mediating pathways, the first step discerns the mediating pathways by determining which exposures influence the mediators and subsequently identifying the mediators that impact the outcome. The discovery process yields a set of exposure-mediators, termed pathways. 

With these pathways fixed, we estimate the average change in the outcome under stochastic shift interventions on exposures, which are further partitioned into direct and indirect effects. We use and extend the methodology first proposed by \cite{diaz_2020}. We use the same efficient influence function for the expected change in outcome given a stochastic shift intervention on the exposure holding the mediator at observed values. We explore numeric integration required for nuisance function estimation and build software for mediation when the exposure is continuous or discrete. The resulting estimates, derived within a cross-validated framework paired with general estimating equations and targeted learning, are asymptotically unbiased with the lowest possible variance, subject to the fulfillment of the unconfoundedness and positivity assumptions. Our proposed method delivers valid confidence intervals, unfettered by the number of exposures, covariates, or the intricacy of the data-generating process, provided the exposures are binned into an arbitrary set of categories. As shown, the numeric integration required for exposures that are modeled truly as continuous induces bias in the estimator which prevents the estimator from converging at the required $\sqrt{n}$ rate, which prohibits our ability to construct valid confidence intervals. 

However, we acknowledge the method's limitations, primarily its requirement for binned exposures and the computational demands of density estimation. Furthermore, interpretation can be challenging in instances where findings are inconsistent. To enhance the reliability and consistency of the data, we recommend reporting the number of folds in which estimates occur and running NOVAPathways with a high number of folds so a majority of data is used for path discovery in each fold.

Notwithstanding these constraints, both our simulations and real-world data applications underscore the robustness and interpretability of our approach, particularly when exposures are binned, which still have valid continuous interpretations. Our NOVAPathways method provides the research community with a statistical machine wherein, the researcher simply puts in a vector of exposures, mediators, covariates, an outcome, estimators used in the Super Learner of each nuisance parameter, and deltas for each respective exposure. The researcher is then provided a table of proportions for each pathway found in the folds and tables providing direct, indirect and total effects for each pathway. 

To support the adoption of semi-parametric methods such as the one we propose, we have made NOVAPathways available via the SuperNOVA R package on GitHub. We believe that by equipping researchers with tools that are not only robust but also flexible, we are inching closer towards solving complex questions in environmental health research. 

\section{Appendix}

\bibliographystyle{unsrtnat}
\bibliography{references}  

\begin{thebibliography}{49}
\providecommand{\natexlab}[1]{#1}
\providecommand{\url}[1]{\texttt{#1}}
\expandafter\ifx\csname urlstyle\endcsname\relax
  \providecommand{\doi}[1]{doi: #1}\else
  \providecommand{\doi}{doi: \begingroup \urlstyle{rm}\Url}\fi

\bibitem[Fedak et~al.(2015)Fedak, Bernal, Capshaw, and Gross]{Fedak2015}
Kristen~M. Fedak, Autumn Bernal, Zachary~A. Capshaw, and Sherilyn Gross.
\newblock {Applying the Bradford Hill criteria in the 21st century: How data
  integration has changed causal inference in molecular epidemiology}.
\newblock \emph{Emerging Themes in Epidemiology}, 12\penalty0 (1):\penalty0
  1--9, 2015.
\newblock ISSN 17427622.
\newblock \doi{10.1186/s12982-015-0037-4}.

\bibitem[Wright(1934)]{wright_1934}
Sewall Wright.
\newblock {The Method of Path Coefficients}.
\newblock \emph{The Annals of Mathematical Statistics}, 5\penalty0
  (3):\penalty0 161 -- 215, 1934.
\newblock \doi{10.1214/aoms/1177732676}.
\newblock URL \url{https://doi.org/10.1214/aoms/1177732676}.

\bibitem[Goldberger(1972)]{goldberg_1972}
Arthur~S Goldberger.
\newblock Structural equation methods in the social sciences.
\newblock \emph{Econometrica: Journal of the Econometric Society}, pages
  979--1001, 1972.

\bibitem[Pearl(2016 - 2016)]{causal_inference_in_stats}
Judea Pearl.
\newblock \emph{Causal inference in statistics : a primer}.
\newblock Wiley, Chichester, West Sussex, 2016 - 2016.
\newblock ISBN 9781119186854.

\bibitem[PEARL(1995)]{pearl_1995}
JUDEA PEARL.
\newblock {Causal diagrams for empirical research}.
\newblock \emph{Biometrika}, 82\penalty0 (4):\penalty0 669--688, 12 1995.
\newblock ISSN 0006-3444.
\newblock \doi{10.1093/biomet/82.4.669}.
\newblock URL \url{https://doi.org/10.1093/biomet/82.4.669}.

\bibitem[Robins(1986)]{robins1986}
James~M Robins.
\newblock A new approach to causal inference in mortality studies with
  sustained exposure periods - application to control of the healthy worker
  survivor effect.
\newblock \emph{Mathematical Modelling}, 7:\penalty0 1393--1512, 1986.

\bibitem[Robins and Greenland(1992{\natexlab{a}})]{robins1992}
James~M Robins and Sander Greenland.
\newblock Identifiability and exchangeability for direct and indirect effects.
\newblock \emph{Epidemiology}, 3\penalty0 (0):\penalty0 143--155,
  1992{\natexlab{a}}.

\bibitem[Rubin(1974)]{rubin1974}
Donald~B Rubin.
\newblock Estimating causal effects of treatments in randomized \&
  nonrandomized studies.
\newblock \emph{Journal of Educational Psychology}, 1974.
\newblock URL \url{http://www.eric.ed.gov/ERICWebPortal/detail?accno=EJ118470}.

\bibitem[Robins and Richardson(2010)]{robins2010}
James~M Robins and Thomas~S Richardson.
\newblock Alternative graphical causal models and the identification of direct
  effects.
\newblock In \emph{Causality and psychopathology: Finding the determinants of
  disorders and their cures}, pages 103--158. 2010.

\bibitem[Robins and Greenland(1992{\natexlab{b}})]{Robins1992IdentifiabilityAE}
James~M. Robins and Sander Greenland.
\newblock Identifiability and exchangeability for direct and indirect effects.
\newblock \emph{Epidemiology}, 3:\penalty0 143--155, 1992{\natexlab{b}}.

\bibitem[Pearl(2001)]{pearl_2001}
Judea Pearl.
\newblock Direct and indirect effects.
\newblock In \emph{Proceedings of the Seventeenth Conference on Uncertainty in
  Artificial Intelligence}, UAI'01, page 411–420, San Francisco, CA, USA,
  2001. Morgan Kaufmann Publishers Inc.
\newblock ISBN 1558608001.

\bibitem[Kennedy(2018)]{kennedy2018}
Edward~H Kennedy.
\newblock Nonparametric causal effects based on incremental propensity score
  interventions.
\newblock \emph{Journal of the American Statistical Association}, pages 1--12,
  2018.

\bibitem[D\'{i}az and van~der Laan(2012)]{diaz2012}
Ivan D\'{i}az and Mark~J van~der Laan.
\newblock Population intervention causal effects based on stochastic
  interventions.
\newblock \emph{Biometrics}, 68\penalty0 (2):\penalty0 541--549, 2012.

\bibitem[Stock(1989)]{stock1989}
James~H Stock.
\newblock Nonparametric policy analysis.
\newblock \emph{Journal of the American Statistical Association}, 84\penalty0
  (406):\penalty0 567--575, 1989.

\bibitem[Robins et~al.(2004)Robins, Hernan, and Siebert]{robins2004}
James~M Robins, Miguel~A Hernan, and Uwe Siebert.
\newblock Effects of multiple interventions.
\newblock \emph{Comparative quantification of health risks: global and regional
  burden of disease attributable to selected major risk factors}, 1:\penalty0
  2191--2230, 2004.

\bibitem[Díaz and Hejazi(2020)]{diaz_2020}
Iván Díaz and Nima~S. Hejazi.
\newblock {Causal Mediation Analysis for Stochastic Interventions}.
\newblock \emph{Journal of the Royal Statistical Society Series B: Statistical
  Methodology}, 82\penalty0 (3):\penalty0 661--683, 02 2020.
\newblock ISSN 1369-7412.
\newblock \doi{10.1111/rssb.12362}.
\newblock URL \url{https://doi.org/10.1111/rssb.12362}.

\bibitem[{Iv{\'{a}}n D{\'{i}}az Mu{\~{n}}oz and Mark van der
  Laan*}(2012)]{diaz_2012}
{Iv{\'{a}}n D{\'{i}}az Mu{\~{n}}oz and Mark van der Laan*}.
\newblock {Population Intervention Causal Effects Based on Stochastic
  Interventions}.
\newblock \emph{Biometrics.}, 68\penalty0 (2):\penalty0 541--549, 2012.
\newblock ISSN 15378276.
\newblock \doi{10.1111/j.1541-0420.2011.01685.x.Population}.
\newblock URL
  \url{https://www.ncbi.nlm.nih.gov/pmc/articles/PMC3624763/pdf/nihms412728.pdf}.

\bibitem[D\'{i}az and van~der Laan(2018)]{diaz2018}
Ivan D\'{i}az and Mark~J van~der Laan.
\newblock Stochastic treatment regimes.
\newblock In \emph{Targeted Learning in Data Science}, pages 219--232.
  Springer, 2018.

\bibitem[Haneuse and Rotnitzky(2013)]{haneuse2013}
Sebastian Haneuse and Andrea Rotnitzky.
\newblock Estimation of the effect of interventions that modify the received
  treatment.
\newblock \emph{Statistics in Medicine}, 2013.

\bibitem[Vansteelandt and VanderWeele(2012)]{vansteelandt2012}
Stijn Vansteelandt and Tyler~J VanderWeele.
\newblock Natural direct and indirect effects on the exposed: effect
  decomposition under weaker assumptions.
\newblock \emph{Biometrics}, 68\penalty0 (4):\penalty0 1019--1027, 2012.

\bibitem[Benkeser and Van Der~Laan(2016)]{hal_paper}
David Benkeser and Mark Van Der~Laan.
\newblock The highly adaptive lasso estimator.
\newblock In \emph{2016 IEEE International Conference on Data Science and
  Advanced Analytics (DSAA)}, pages 689--696, 2016.
\newblock \doi{10.1109/DSAA.2016.93}.

\bibitem[{Milborrow. Derived from mda:mars by T. Hastie and R.
  Tibshirani.}(2011)]{earth}
S.~{Milborrow. Derived from mda:mars by T. Hastie and R. Tibshirani.}
\newblock \emph{earth: Multivariate Adaptive Regression Splines}, 2011.
\newblock URL \url{http://CRAN.R-project.org/package=earth}.
\newblock R package.

\bibitem[Ripley and Venables(2021)]{polspline}
B.~D. Ripley and W.~Venables.
\newblock \emph{polspline: Polynomial Spline Routines}, 2021.
\newblock URL \url{https://CRAN.R-project.org/package=polspline}.
\newblock R package version 1.1.26.

\bibitem[Coyle et~al.(2022)Coyle, Hejazi, Phillips, {van der Laan}, and {van
  der Laan}]{hal9001}
Jeremy~R Coyle, Nima~S Hejazi, Rachael~V Phillips, Lars~WP {van der Laan}, and
  Mark~J {van der Laan}.
\newblock \emph{{hal9001}: The scalable highly adaptive lasso}, 2022.
\newblock URL \url{https://github.com/tlverse/hal9001}.
\newblock R package version 0.4.3.

\bibitem[McCoy et~al.(2023)McCoy, Hubbard, and der Laan]{McCoy2023}
David McCoy, Alan Hubbard, and Mark~Van der Laan.
\newblock Cvtreemle: Efficient estimation of mixed exposures using data
  adaptive decision trees and cross-validated targeted maximum likelihood
  estimation in r.
\newblock \emph{Journal of Open Source Software}, 8\penalty0 (82):\penalty0
  4181, 2023.
\newblock \doi{10.21105/joss.04181}.
\newblock URL \url{https://doi.org/10.21105/joss.04181}.

\bibitem[Hubbard et~al.(2016)Hubbard, Kherad-Pajouh, and {Van Der
  Laan}]{hubbard2016}
Alan~E. Hubbard, Sara Kherad-Pajouh, and Mark~J. {Van Der Laan}.
\newblock {Statistical Inference for Data Adaptive Target Parameters}.
\newblock \emph{International Journal of Biostatistics}, 12\penalty0
  (1):\penalty0 3--19, 2016.
\newblock ISSN 15574679.
\newblock \doi{10.1515/ijb-2015-0013}.

\bibitem[Hejazi et~al.(2022)Hejazi, Benkeser, and {van der Laan}]{haldensify}
Nima~S Hejazi, David Benkeser, and Mark~J {van der Laan}.
\newblock \emph{{haldensify}: Highly adaptive lasso conditional density
  estimation}, 2022.
\newblock URL \url{https://github.com/nhejazi/haldensify}.
\newblock R package version 0.2.3.

\bibitem[Marschner(2011)]{glm}
Ian~C. Marschner.
\newblock {glm2}: Fitting generalized linear models with convergence problems.
\newblock \emph{The R Journal}, 3:\penalty0 12--15, 2011.

\bibitem[Friedman et~al.(2010)Friedman, Hastie, and Tibshirani]{elasticnet}
Jerome Friedman, Trevor Hastie, and Robert Tibshirani.
\newblock Regularization paths for generalized linear models via coordinate
  descent.
\newblock \emph{Journal of Statistical Software}, 33\penalty0 (1):\penalty0
  1--22, 2010.
\newblock \doi{10.18637/jss.v033.i01}.
\newblock URL \url{https://www.jstatsoft.org/v33/i01/}.

\bibitem[Wright and Ziegler(2017)]{ranger}
Marvin~N. Wright and Andreas Ziegler.
\newblock {ranger}: A fast implementation of random forests for high
  dimensional data in {C++} and {R}.
\newblock \emph{Journal of Statistical Software}, 77\penalty0 (1):\penalty0
  1--17, 2017.
\newblock \doi{10.18637/jss.v077.i01}.

\bibitem[Chen and Guestrin(2016)]{xgboost}
Tianqi Chen and Carlos Guestrin.
\newblock {XGBoost}: A scalable tree boosting system.
\newblock In \emph{Proceedings of the 22nd ACM SIGKDD International Conference
  on Knowledge Discovery and Data Mining}, KDD '16, pages 785--794, New York,
  NY, USA, 2016. ACM.
\newblock ISBN 978-1-4503-4232-2.
\newblock \doi{10.1145/2939672.2939785}.
\newblock URL \url{http://doi.acm.org/10.1145/2939672.2939785}.

\bibitem[van~der Laan~Mark et~al.(2007)van~der Laan~Mark, C, and E.]{SL_2008}
J.~van~der Laan~Mark, Polley~Eric C, and Hubbard~Alan E.
\newblock Super learner.
\newblock \emph{Statistical Applications in Genetics and Molecular Biology},
  6\penalty0 (1):\penalty0 1--23, 2007.
\newblock URL
  \url{https://EconPapers.repec.org/RePEc:bpj:sagmbi:v:6:y:2007:i:1:n:25}.

\bibitem[Zipf et~al.(2013)Zipf, Chiappa, Porter, Ostchega, Lewis, and
  Dostal]{zipf2013national}
George Zipf, Michele Chiappa, Kathryn~S Porter, Yechiam Ostchega, Brenda~G
  Lewis, and Jennifer Dostal.
\newblock National health and nutrition examination survey: Plan and
  operations, 1999–2010.
\newblock \emph{Vital and health statistics. Series 1, Programs and collection
  procedures}, 56:\penalty0 1--37, 2013.

\bibitem[Wilson et~al.(2017)Wilson, Bryan, Cranston, Kitzes, Nederbragt, and
  Teal]{wilson2017good}
Greg Wilson, Jennifer Bryan, Karen Cranston, Justin Kitzes, Lex Nederbragt, and
  Tracy~K Teal.
\newblock Good enough practices in scientific computing.
\newblock \emph{PLoS computational biology}, 13\penalty0 (6), 2017.

\bibitem[Hernan and Robins(2010)]{hernan2010causal}
Miguel~A Hernan and James~M Robins.
\newblock \emph{Causal Inference: What If}.
\newblock Chapman \& Hall/CRC, 2010.

\bibitem[Balali-Mood et~al.(2021)Balali-Mood, Naseri, Tahergorabi, Khazdair,
  and Sadeghi]{Balali_Mood2021}
Mahdi Balali-Mood, Kobra Naseri, Zoya Tahergorabi, Mohammad~Reza Khazdair, and
  Mahmood Sadeghi.
\newblock {Toxic Mechanisms of Five Heavy Metals: Mercury, Lead, Chromium,
  Cadmium, and Arsenic}.
\newblock \emph{Frontiers in Pharmacology}, 12\penalty0 (April):\penalty0
  1--19, 2021.
\newblock ISSN 16639812.
\newblock \doi{10.3389/fphar.2021.643972}.

\bibitem[Horton and Kleinman(2007)]{horton2007much}
Nicholas~J Horton and Ken~P Kleinman.
\newblock Much ado about nothing: A comparison of missing data methods and
  software to fit incomplete data regression models.
\newblock \emph{The American Statistician}, 61\penalty0 (1):\penalty0 79--90,
  2007.

\bibitem[Grzela et~al.(2017)Grzela, Litwiniuk, Krejner, and
  Grzela]{grzela2017oxidative}
Katarzyna Grzela, Malgorzata Litwiniuk, Alicja Krejner, and Tomasz Grzela.
\newblock Oxidative stress and bronchial asthma in children—causes or
  consequences?
\newblock \emph{Frontiers in Pediatrics}, 5:\penalty0 129, 2017.
\newblock \doi{10.3389/fped.2017.00129}.
\newblock URL \url{https://www.ncbi.nlm.nih.gov/pmc/articles/PMC5523023/}.

\bibitem[Vargas et~al.(2021)Vargas, Buendía, Alean, Castillo, Gomez, Parra,
  Ramos-Bonilla, and González-García]{vargas2021metals}
Daniela~Vargas Vargas, Jorge A.~Buendía Buendía, Diego A.~Alean Alean, María
  Juliana~Castillo Castillo, Magda E.~Gomez Gomez, Edgar J.~Parra Parra, Juan
  P. Ramos-Bonilla Ramos-Bonilla, and Mauricio González-García
  González-García.
\newblock Metals and metalloids in asthma: A role for environmental exposure?
\newblock \emph{Frontiers in Pharmacology}, 12:\penalty0 643972, 2021.
\newblock \doi{10.3389/fphar.2021.643972}.
\newblock URL
  \url{https://www.frontiersin.org/articles/10.3389/fphar.2021.643972/full}.

\bibitem[Esmaeilzadeh et~al.(2021)Esmaeilzadeh, Ahmadi, Jafari, Farajollahi,
  Hekmatnia, Jafari, and Kalantari]{esmaeilzadeh2021role}
Abbas Esmaeilzadeh, Ali Ahmadi, Sayed~Jalal Jafari, Mohammad~Mahdi Farajollahi,
  Fatemeh Hekmatnia, Afshin Jafari, and Kioomars Kalantari.
\newblock Role of oxidative stress in respiratory diseases: from molecular
  mechanisms to therapeutic approaches.
\newblock \emph{Respiratory Research}, 22:\penalty0 210, 2021.
\newblock \doi{10.1186/s12931-021-01801-4}.
\newblock URL \url{https://www.ncbi.nlm.nih.gov/pmc/articles/PMC8330548/}.

\bibitem[Kong et~al.(2023)Kong, Liu, and Olatunji]{Kong2023}
Zhiyang Kong, Chunhong Liu, and Opeyemi~Joshua Olatunji.
\newblock {Asperuloside attenuates cadmium-induced toxicity by inhibiting
  oxidative stress, inflammation, fibrosis and apoptosis in rats}.
\newblock \emph{Scientific Reports}, 13\penalty0 (1):\penalty0 5698, 2023.
\newblock ISSN 2045-2322.
\newblock \doi{10.1038/s41598-023-29504-0}.
\newblock URL \url{https://doi.org/10.1038/s41598-023-29504-0}.

\bibitem[Rahman and Adcock(2006)]{Rahman2006}
Irfan Rahman and I.~M. Adcock.
\newblock {Oxidative stress and redox regulation of lung inflammation in COPD}.
\newblock \emph{European Respiratory Journal}, 28\penalty0 (1):\penalty0
  219--242, 2006.
\newblock ISSN 09031936.
\newblock \doi{10.1183/09031936.06.00053805}.

\bibitem[Hejazi and D{\'\i}az(2020)]{hejazi2020medshift}
Nima~S Hejazi and Iv{\'a}n D{\'\i}az.
\newblock \emph{{medshift}: Causal mediation analysis for stochastic
  interventions}, 2020.
\newblock URL \url{https://github.com/nhejazi/medshift}.
\newblock R package version 0.1.4.

\bibitem[Díaz et~al.(2021)Díaz, Williams, Hoffman, and Schneck]{lmtp_article}
Iván Díaz, Nicholas Williams, Katherine Hoffman, and Edward Schneck.
\newblock Non-parametric causal effects based on longitudinal modified
  treatment policies.
\newblock \emph{Journal of the American Statistical Association}, 2021.
\newblock \doi{10.1080/01621459.2021.1955691}.

\bibitem[Williams and Díaz(2020)]{lmtp_manual}
Nicholas Williams and Iván Díaz.
\newblock \emph{lmtp: {Non}-parametric {Causal} {Effects} of {Feasible}
  {Interventions} {Based} on {Modified} {Treatment} {Policies}}, 2020.
\newblock URL \url{https://github.com/nt-williams/lmtp}.
\newblock R package version 1.3.1.

\bibitem[Coyle et~al.(2018)Coyle, Hejazi, Malenica, and Sofrygin]{coyle2018}
Jeremy~R Coyle, Nima~S Hejazi, Ivana Malenica, and Oleg Sofrygin.
\newblock sl3: Modern pipelines for machine learning and super learning.
\newblock \url{https://github.com/tlverse/sl3}, 2018.
\newblock R package version 1.1.0.

\bibitem[Steen et~al.(2017)Steen, Loeys, Moerkerke, and
  Vansteelandt]{Steen2017}
J~Steen, T~Loeys, B~Moerkerke, and S~Vansteelandt.
\newblock medflex: An r package for flexible mediation analysis using natural
  effect models.
\newblock \emph{Journal of Statistical Software}, 76:\penalty0 1--46, 2017.
\newblock \doi{10.18637/jss.v076.i11}.

\bibitem[Tingley et~al.(2014)Tingley, Yamamoto, Hirose, Keele, and
  Imai]{mediation_package}
Dustin Tingley, Teppei Yamamoto, Kentaro Hirose, Luke Keele, and Kosuke Imai.
\newblock {mediation}: {R} package for causal mediation analysis.
\newblock \emph{Journal of Statistical Software}, 59\penalty0 (5):\penalty0
  1--38, 2014.
\newblock URL \url{http://www.jstatsoft.org/v59/i05/}.

\bibitem[Vaughan and Dancho(2022)]{furrr_2022}
Davis Vaughan and Matt Dancho.
\newblock \emph{furrr: Apply Mapping Functions in Parallel using Futures},
  2022.
\newblock https://github.com/DavisVaughan/furrr,
  https://furrr.futureverse.org/.

\end{thebibliography}

\end{document}